\documentclass[referee]{raa}           
\usepackage[compatibility=false]{caption}
\usepackage{graphicx,times}
\usepackage{natbib}
\usepackage{amssymb,amsmath}
\bibpunct{(}{)}{;}{a}{}{,}

\usepackage{multirow}  
\usepackage{array} 

\usepackage{booktabs}  
\usepackage{makecell}  
\usepackage{graphicx}  

\usepackage{caption}  
\usepackage{placeins} 
\usepackage[pagebackref=true]{hyperref}

\begin{document}

\title{Segmentation and Celestial Mapping of Unobservable Regions in Nighttime All-sky Images for the Mephisto Observations}

\volnopage{ {\bf 20XX} Vol.\ {\bf X} No. {\bf XX}, 000--000}
\setcounter{page}{1}

   \author{Jian Cui 
   \inst{1}, Guo-Wang Du\inst{1,2}\thanks{email: dugking@ynu.edu.cn}, Xin-Zhong Er\inst{3}\thanks{email: phioen@163.com}, Chu-Xiang Li\inst{3}, Jun-Fan Hou\inst{3}, Yu-Xin Xin\inst{4,5}, Xiang-kun Liu\inst{1,2}, Xiao-Wei Liu\inst{1,2}
   }

   \institute{South-Western Institute for Astronomy Research, Yunnan University, Kunming, Yunnan 650504, China\\
        \and
            Key Laboratory of Survey Science of Yunnan Province, Yunnan University, Kunming, Yunnan 650500, People’s Republic of China\\
        \and
             Tianjin Astrophysics Center, Tianjin Normal University, Tianjin, 300387, China\\
	\and
Yunnan Observatories, Chinese Academy of Sciences, Kunming 650011, China;\\
\and 
 Key Laboratory for the Structure and Evolution of Celestial Objects, Yunnan Observatories, Kunming 650011, China\\
\vs \no
   {\small Received 20XX Month Day; accepted 20XX Month Day}
}

\abstract{Accurate identification of unobservable regions in nighttime is essential for autonomous scheduling and data quality control in observations. Traditional methods—such as infrared sensing or photometric extinction—provide only coarse, non-spatial estimates of sky clarity, making them insufficient for real-time decision-making. This not only wastes observing time but also introduces contamination when telescopes are directed toward cloud-covered or moonlight-affected regions. To address these limitations, we propose a deep learning-based segmentation framework that provides pixel-level masks of unobservable areas using all-sky images. Supported by a manually annotated dataset of nighttime images, our method enables precise detection of cloud- and moonlight-affected regions. The segmentation results are further mapped to celestial coordinates through Zenithal Equal-Area projection, allowing seamless integration with observation control systems (OCS) for real-time cloud-aware scheduling. While developed for the Mephisto telescope, the framework is generalizable and applicable to other wide-field robotic observatories equipped with all-sky monitoring.   
}

    \authorrunning{J. Cui et al.}            
    \titlerunning{Segmentation and Celestial Mapping of Unobservable Regions}    
    \title{Segmentation and Celestial Mapping of Unobservable Regions in Nighttime All-sky Images for the Mephisto Observations}  
    \maketitle

\section{Introduction}  
\label{sect:intro}  
The rapid development of time-domain astronomy has been driven by the advent of wide-field, high-cadence sky surveys, A large number of transients and varibales have been detected and studied in details. In response to these scientific demands, a new generation of large-scale survey telescopes has emerged. Notable examples include the Zwicky Transient Facility (ZTF), which uses a panoramic camera for high cadence wide-area optical monitoring (\citealt{Dekany+etal+2020}); the Wide Field Survey Telescope (WFST, “Mozi”) at Saishiteng Mountain Observatory in Qinghai, China (\citealt{Wang+etal+2023}); and the Multichannel Photometric Survey Telescope (Mephisto) at Yunnan University’s Lijiang Observatory, which allows simultaneous multiband imaging, i.e., $u/g/i$ or $v/r/z$ (\citealt{Yuan+etal+2020}). The operational systems are designed for robotic operation, rapid follow-up observations automatically, and continuous data acquisition.  

Space-borne missions, such as the Einstein Probe (EP) and the multi-band astronomical Variable Objects Monitor (SVOM), are powerful tools for detecting high-energy transients and variable sources (\citealt{Cordier+etal+2018, Yuan+etal+2025}). However, only with the support of rapid multi-band observations can we study the mechanisms of these events (\citealt{zhang+etal+2025, Liu+etal+2025, Yin+etal+2025, Du+etal+2025}) . While satellite observations are immune to terrestrial weather, the immediate follow-up observations by ground-based telescopes rely on weather conditions. Consequently, real-time monitoring of atmospheric conditions is essential to maximize observing efficiency and minimize missed scientific opportunities, especially for robotic infrastructure. Environmental awareness—particularly the detection of clouds and scattered moonlight—is crucial for maintaining data quality and optimizing telescope scheduling.

Atmospheric effects, primarily cloud cover and scattered moonlight, cause regions of the sky that are unobservable and can severely degrade photometric measurements in automated surveys. These conditions obscure sources, distort backgrounds and point spread functions, and undermine both data quality and calibration. Early identification and avoidance of such regions are therefore vital for preserving survey efficiency and scientific integrity.  

Different techniques have been developed and deployed to assess nighttime sky observability. Infrared cloud sensors (IRCS) utilize thermal contrasts to infer cloudiness and provide rapid, observatory-wide assessments (\citealt{Takato+etal+2003, Lewis+etal+2010, Redman+etal+2018, xin+etal+2020}). 
Ground based sky imagers (SIs), originally developed in atmospheric science, enable real-time cloud mapping with visible and infrared cameras, aided by robust image-processing pipelines (\citealt{ Cazorla+etal+2008, Ghonima+etal+2012, Shields+etal+2013}).
By comparing observed stellar photometry with catalog, one can estimate the atmospheric transmission spatially, and schedule the observation accordingly (\citealt{Ofek+etal+2012, Chambers+etal+2016, Tonry+etal+2018}). 
Classical image-processing techniques—including thresholding, histogram equalization, and morphological filtering—offer computational efficiency and have been broadly adopted for cloud segmentation tasks (\citealt{Souza-Echer+etal+2006,  Calbo+Sabburg+2008, Kazantzidis+etal+2012, Taravat+etal+2015}).
While these approaches have enabled significant progress and remain valuable baselines, several key challenges persist, including precise segmentation at the pixel level, robustness under varying illumination or complex backgrounds, and timely response for modern automated observatory control. Addressing these limitations motivates further exploration of advanced methods with operational efficiency in challenging nighttime conditions.

These challenges underscore the need for data-driven frameworks capable of robust, pixel-level segmentation of unobservable regions in nighttime all-sky images. Recent advances in machine learning—particularly convolutional neural networks (CNNs)—have demonstrated great potential in astronomical applications such as transient detection and image analysis (\citealt{Duev+etal+2019, Muthukrishna+etal+2019, Hausen+etal+2020, Forster+2021, Fu+etal+2024, Jia+etal+2023, Sun+etal+2023}). Semantic segmentation models like U-Net have proven effective in delivering pixel-level discrimination in complex nighttime sky imagery, consistently outperforming threshold-based or empirical methods (\citealt{Li+etal+2012, Liu+etal+2015, Cheng+Lin+2017, Xie+etal+2020, Fabel+etal+2022}). Nevertheless, several challenges remain: complex cloud morphology, weak illumination, and interference from stars or artificial light complicate both training and annotation, while hardware constraints limit real-time deployment in remote observatories. Furthermore, segmentation outputs in image coordinates are not directly compatible with observatory control systems (OCS), hindering their seamless integration into observation planning.  

The Mephisto project, initiated by Yunnan University in 2018, focuses on high-cadence, wide-field, and autonomous time-domain surveys (\citealt{Yuan+etal+2020}). However, in its current phase of unattended operation, Mephisto’s dynamic scheduling and robust target execution are frequently hindered by nighttime atmospheric effects—especially cloud cover and scattered moonlight—that degrade photometric quality but often escape detection by conventional monitors.  

In this work, we propose a deep learning-based framework tailored for unattended, wide-field telescopes to achieve robust, real-time segmentation of unobservable sky regions in all-sky images. Our approach leverages recent advances in semantic segmentation and incorporates a coordinate transformation pipeline, enabling seamless integration with observatory control systems. Although developed and validated on Mephisto, the framework is broadly applicable to other survey facilities equipped with all-sky cameras, providing a scalable and practical foundation for visibility-aware scheduling and fully autonomous operations in time-domain astronomy. Section~\ref{sec:data} describes the all-sky camera setup, data preprocessing, and annotation protocol at Lijiang Observatory. Section~\ref{sec:methodology} details our segmentation framework, including the improved U-Net model and its training. Section~\ref{sec:coordinate_mapping} outlines the procedure for mapping segmentation results to celestial coordinates for operational scheduling. Finally, Section~\ref{sec:conclusion} summarizes our main result and discusses future plans. 

\section{All-Sky Camera System and Nighttime Image Dataset}  
\label{sec:data}  

The all-sky camera facility is installed at Lijiang Observatory (IAU code: 044), Yunnan Astronomical Observatories, Chinese Academy of Sciences (CAS), situated at $100^\circ 01' 48''$ east longitude,  $26^\circ 41' 42''$ north latitude, and an altitude of approximately 3,200 meters (\citealt{Wang+etal+2019}). The site benefits from minimal air pollution and low dust content, resulting in a clean and transparent atmosphere that provides excellent conditions for astronomical observations (\citealt{Tan+etal+2002}). It is equipped with a high-sensitivity, low-noise CMOS sensor paired with a fisheye lens, providing a $360^\circ$ field of view for comprehensive sky coverage. It is capable of capturing high-resolution images in real time—even under extremely low-light nighttime conditions—and supports high-frequency image acquisition as well as real-time data transmission. These features make the facility particularly effective for long-term sky survey operations. While the system enables continuous, all-weather monitoring of the entire sky, regular maintenance and human supervision are still necessary to ensure stable performance and reliable data quality (\citealt{xin+etal+2020}).

\begin{figure}[t]  
    \centering  
    \begin{minipage}[b]{0.23\linewidth}  
        \centering  
        \includegraphics[width=\linewidth, height=\linewidth]{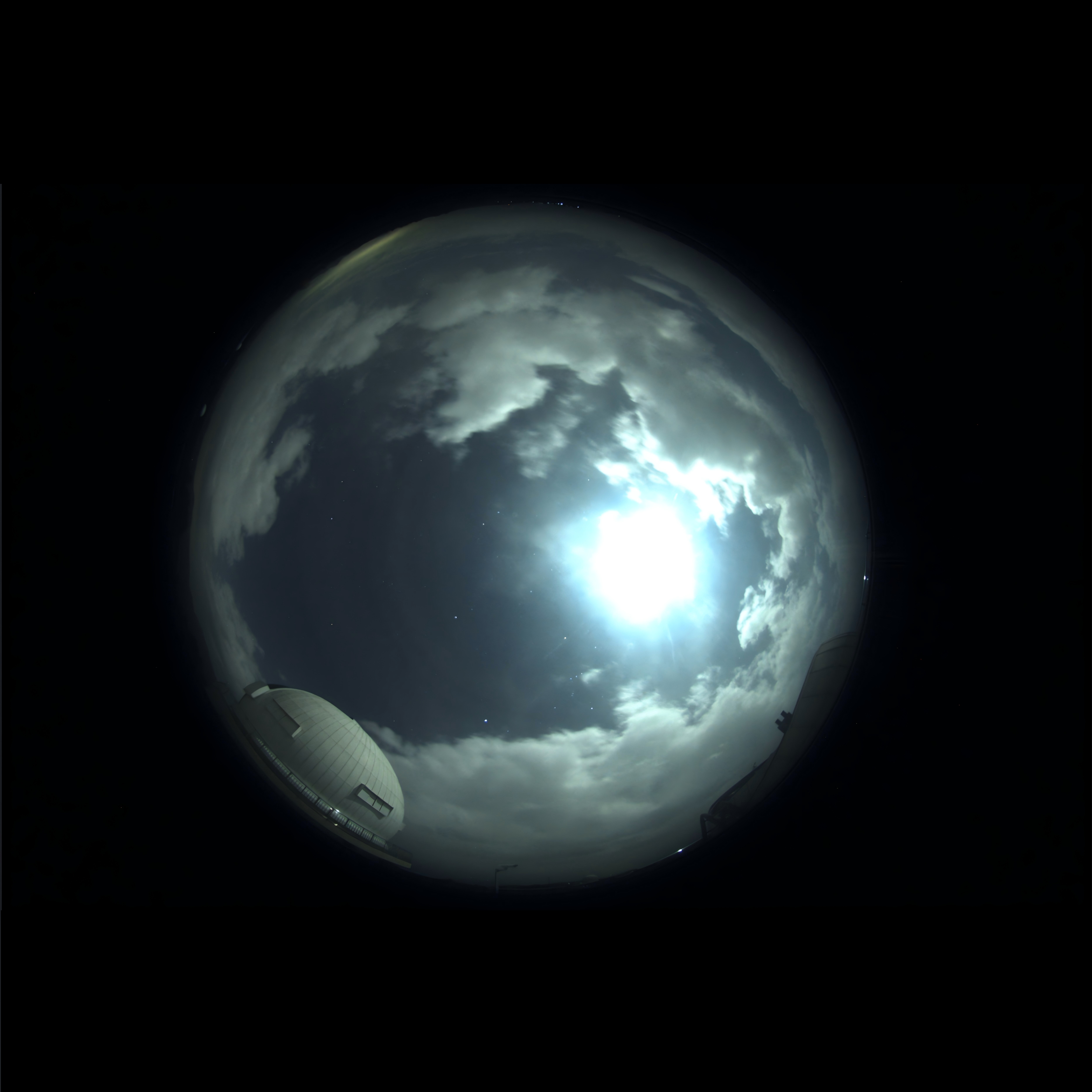} \\
        \small (a)  
    \end{minipage}  
    \begin{minipage}[b]{0.23\linewidth}  
        \centering  
        \includegraphics[width=\linewidth, height=\linewidth]{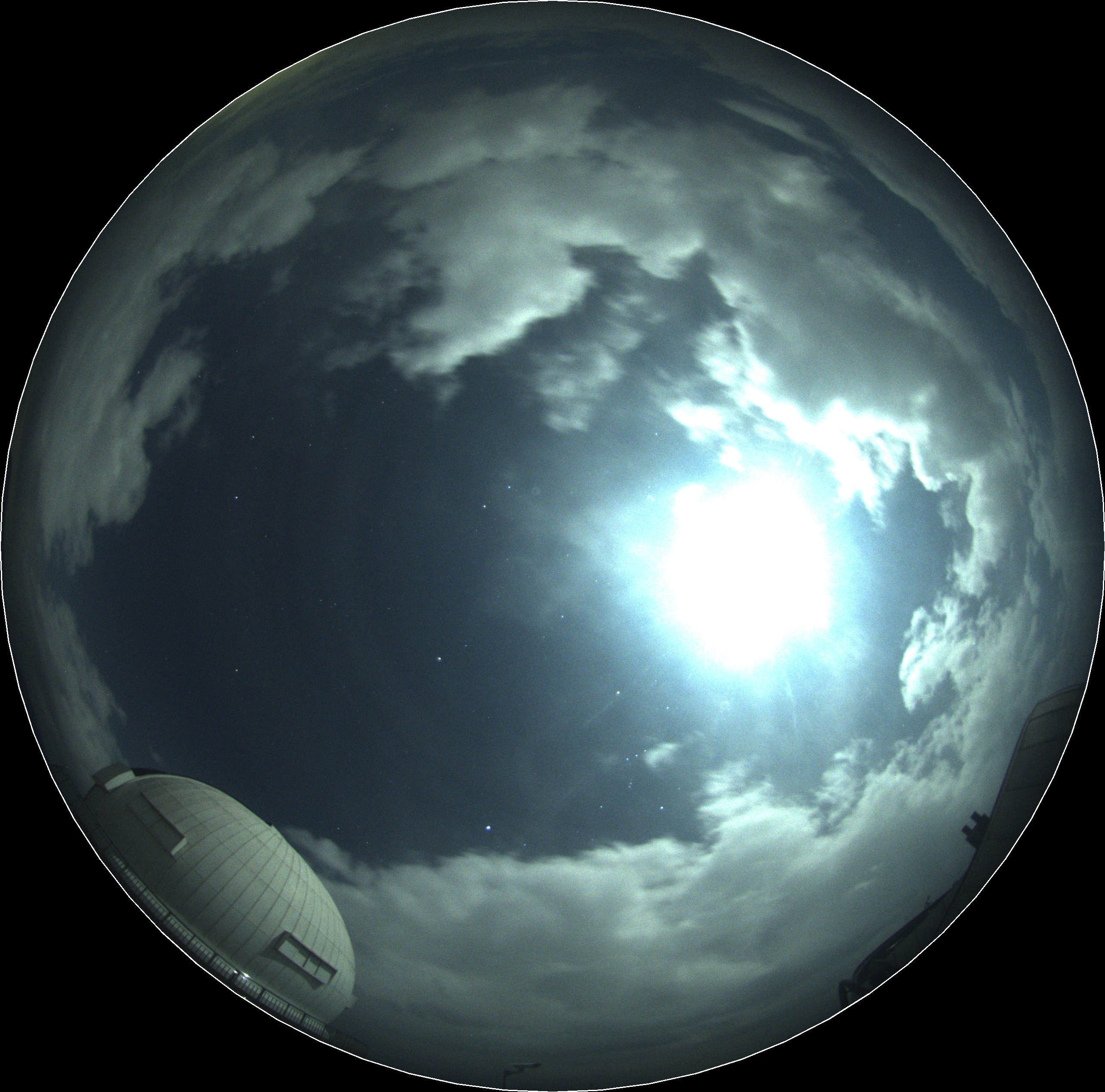} \\
        \small (b)  
    \end{minipage}  
    \begin{minipage}[b]{0.23\linewidth}  
        \centering  
        \includegraphics[width=\linewidth, height=\linewidth]{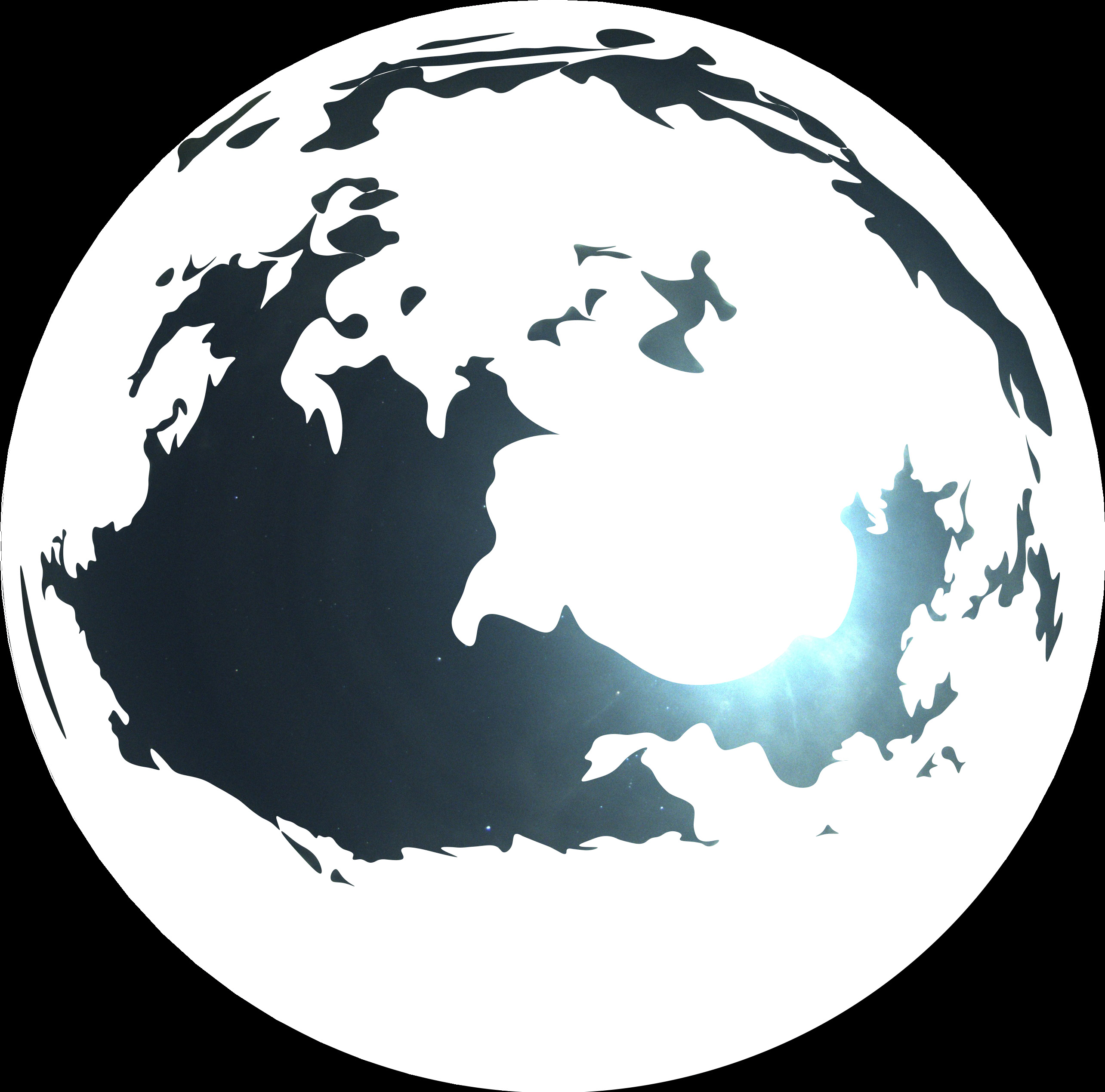} \\
        \small (c)  
    \end{minipage}  
    \begin{minipage}[b]{0.23\linewidth}  
        \centering  
        \includegraphics[width=\linewidth, height=\linewidth]{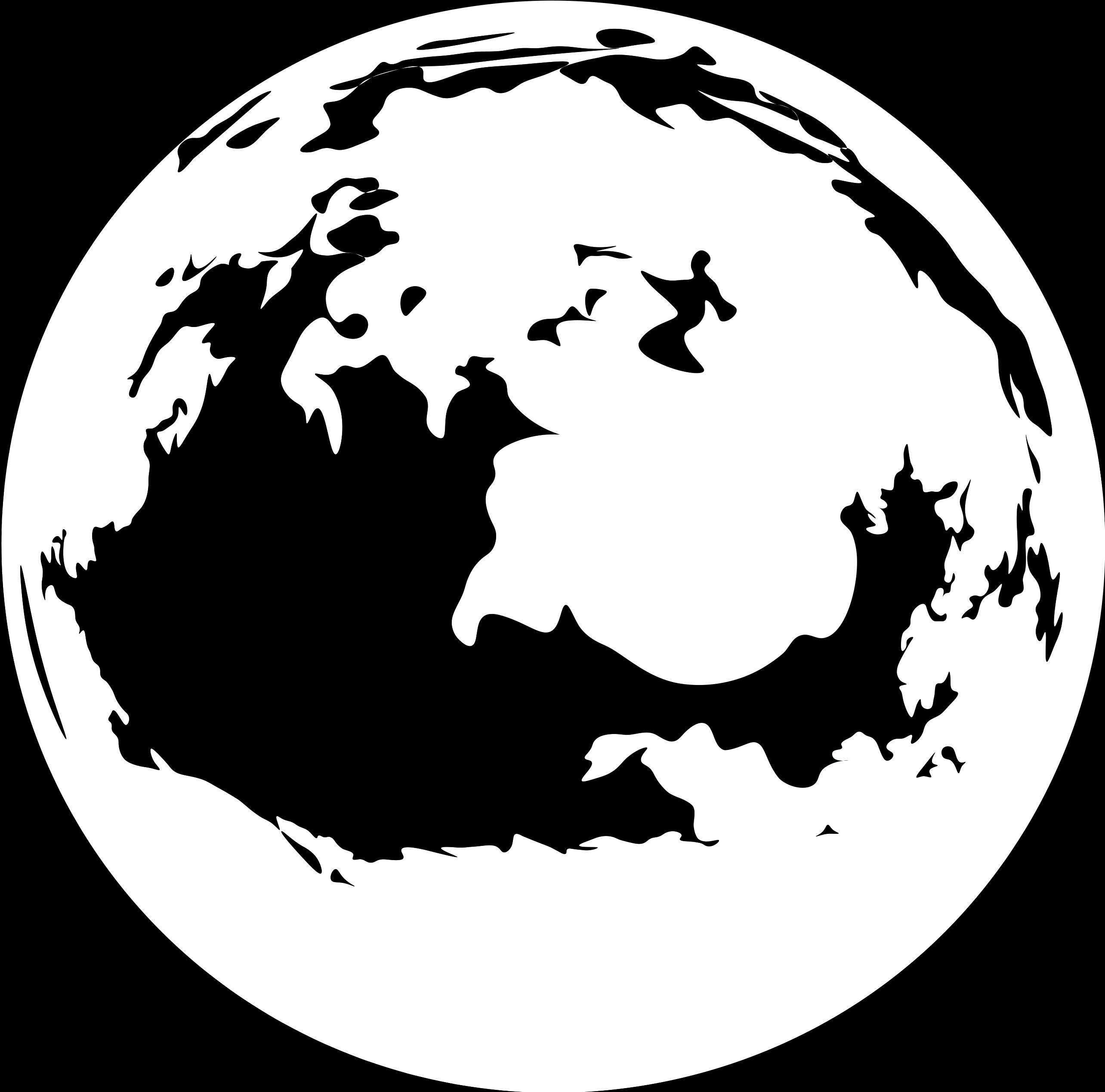} \\
        \small (d)  
    \end{minipage}  
    \caption{Image preprocessing workflow: (a) Raw image captured by the all-sky camera; (b) Cropped and resized circular region; (c) Annotated image with masked obstructions; (d) Binarized mask highlighting cloud objects (white) and background (black). }  
    \label{fig:images}  
\end{figure}

To construct a standardized nighttime sky dataset, 2,000 representative images are systematically selected from the all-sky camera data collected between January and September 2024, specifically during nighttime hours (\texttt{20:00} to \texttt{04:00}). This dataset covers a comprehensive array of nighttime weather conditions observable by the all-sky camera, including clear skies, thin or scattered clouds, partially cloudy conditions, and fully overcast skies. By encompassing this diversity of atmospheric phenomena captured during nighttime observations, the dataset ensures both temporal variability and meteorological representativeness. The raw images, as illustrated in Fig.~\ref{fig:images}(a), initially exhibit variations in dimensions (4144 × 2822 pixels) in 24-bit RGB color format. To ensure consistency and maintain high data quality, a dedicated preprocessing workflow is applied.  

\begin{enumerate}
    \item \textbf{Image Cropping and Resizing:} Each image is cropped to the largest inscribed circle containing cloud features and then resized to 2568$\times$2568 pixels, thereby preserving essential sky and cloud structures while standardizing the input size. Fig.~\ref{fig:images}(b) shows the cropped and resized circular region.  
    
    \item \textbf{Annotation:} The cropped images are manually annotated using image editing software, where obstructed or irrelevant regions are masked to generate precise ground-truth labels. During annotation, static regions such as the ground, observatory dome, and DIMM tower are masked as “unobservable regions” to maintain a consistent data format. Since these structures occupy fixed positions and do not obstruct actual telescope observations, they are excluded from subsequent analysis by applying a static mask template. As a result, only dynamic atmospheric factors—such as clouds and moonlight—are considered in visibility assessment and observation scheduling. Fig.~\ref{fig:images}(c) displays the annotated image with masked obstructions.

    \item \textbf{Binarization:} The annotated masks are batch-processed to produce standardized binary images: cloud regions are represented in white (pixel value 255), and the background is black (pixel value 0). These binary masks serve as crucial inputs for subsequent machine learning analyses. Fig.~\ref{fig:images}(d) depicts the corresponding binarized mask, where cloud regions appear in white and the background in black.

\end{enumerate}

Each image in the dataset is further categorized according to the proportion of unobservable regions—areas obscured by heavy cloud cover or other obstructions that hinder the detection of celestial objects. Three categories are defined: High, Moderate, and Low Unobservable Regions, as summarized in Table~\ref{table:classification}. Notably, the Low Unobservable Region category includes a fixed 2\% allocation due to the presence of buildings in certain images. In the adjusted 2,000-image dataset, Low Unobservable Region images account for 34.02\%, Moderate for 29.77\%, and High for 36.21\%. This systematic categorization facilitates detailed analysis of the impact of cloud coverage on image processing and segmentation, thereby supporting the development and evaluation of robust nighttime sky analysis algorithms.  

\begin{figure*}[t]  
    \centering  
    \renewcommand{\arraystretch}{1.2}  
    \setlength{\tabcolsep}{1pt}  
    \begin{tabular}{  
        >{\centering\arraybackslash}m{1.6cm}  
        *{6}{>{\centering\arraybackslash}m{1.9cm}}  
    }  
        \toprule  
        \textbf{Category} &  
        \multicolumn{2}{c}{\makecell[c]{\textbf{High}\\\textbf{Unobservable}\\\textbf{Region}}} &  
        \multicolumn{2}{c}{\makecell[c]{\textbf{Moderate}\\\textbf{Unobservable}\\\textbf{Region}}} &  
        \multicolumn{2}{c}{\makecell[c]{\textbf{Low}\\\textbf{Unobservable}\\\textbf{Region}}} \\
        \midrule  
        \makecell[c]{\textbf{Raw}\\\textbf{Images}} &  
        \includegraphics[width=1.8cm, height=1.8cm, keepaspectratio]{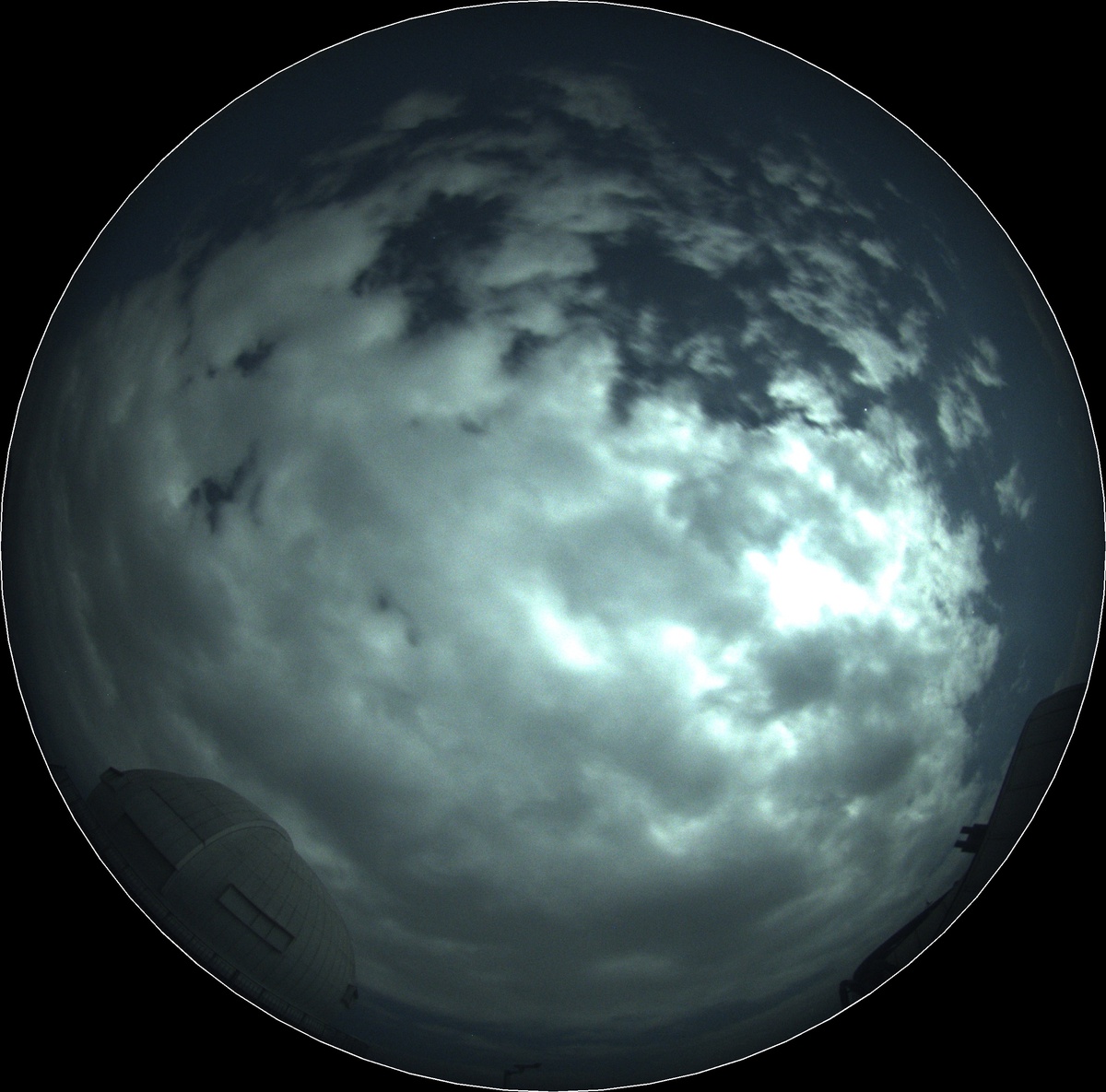} &  
        \includegraphics[width=1.8cm, height=1.8cm, keepaspectratio]{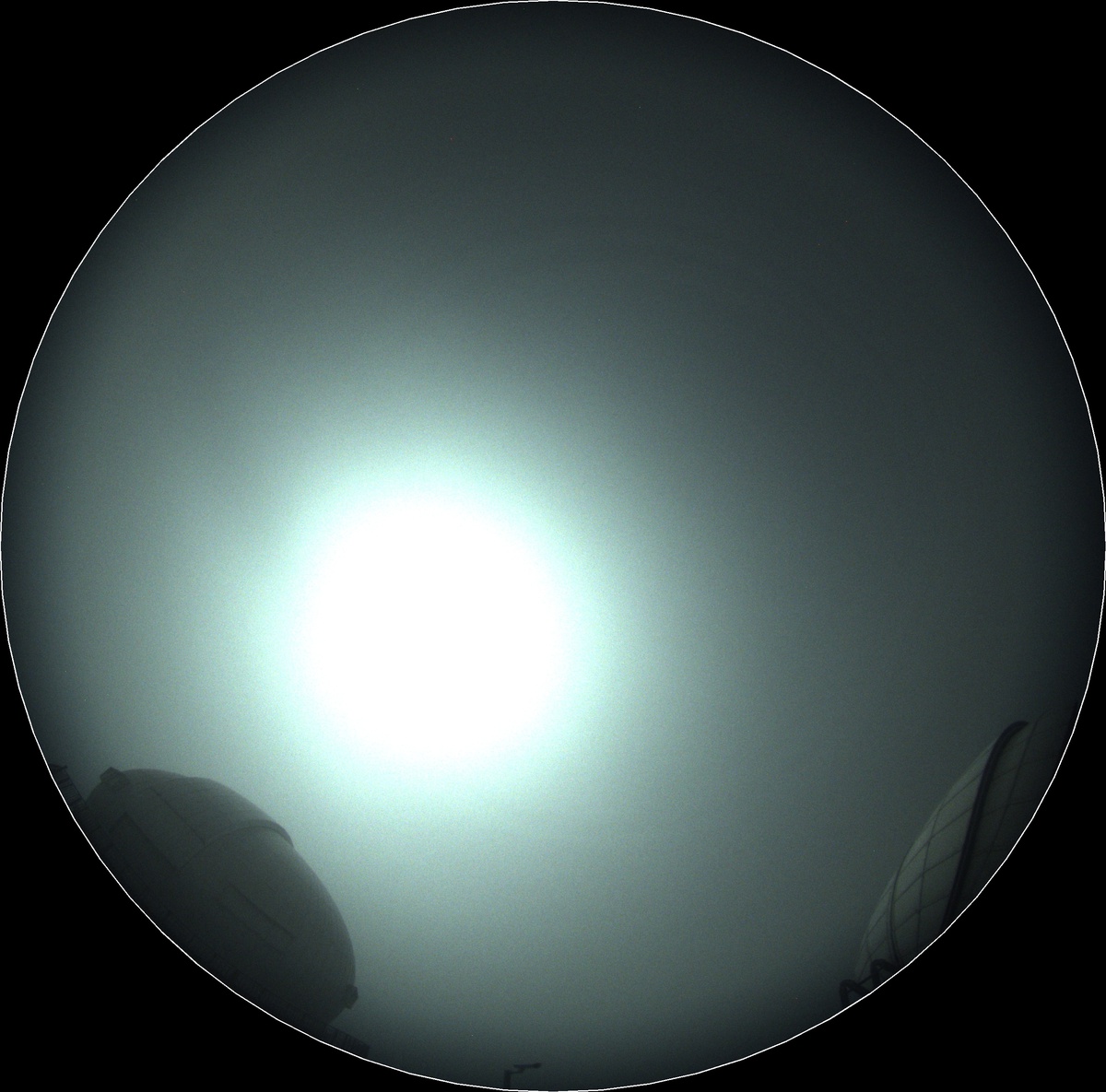} &  
        \includegraphics[width=1.8cm, height=1.8cm, keepaspectratio]{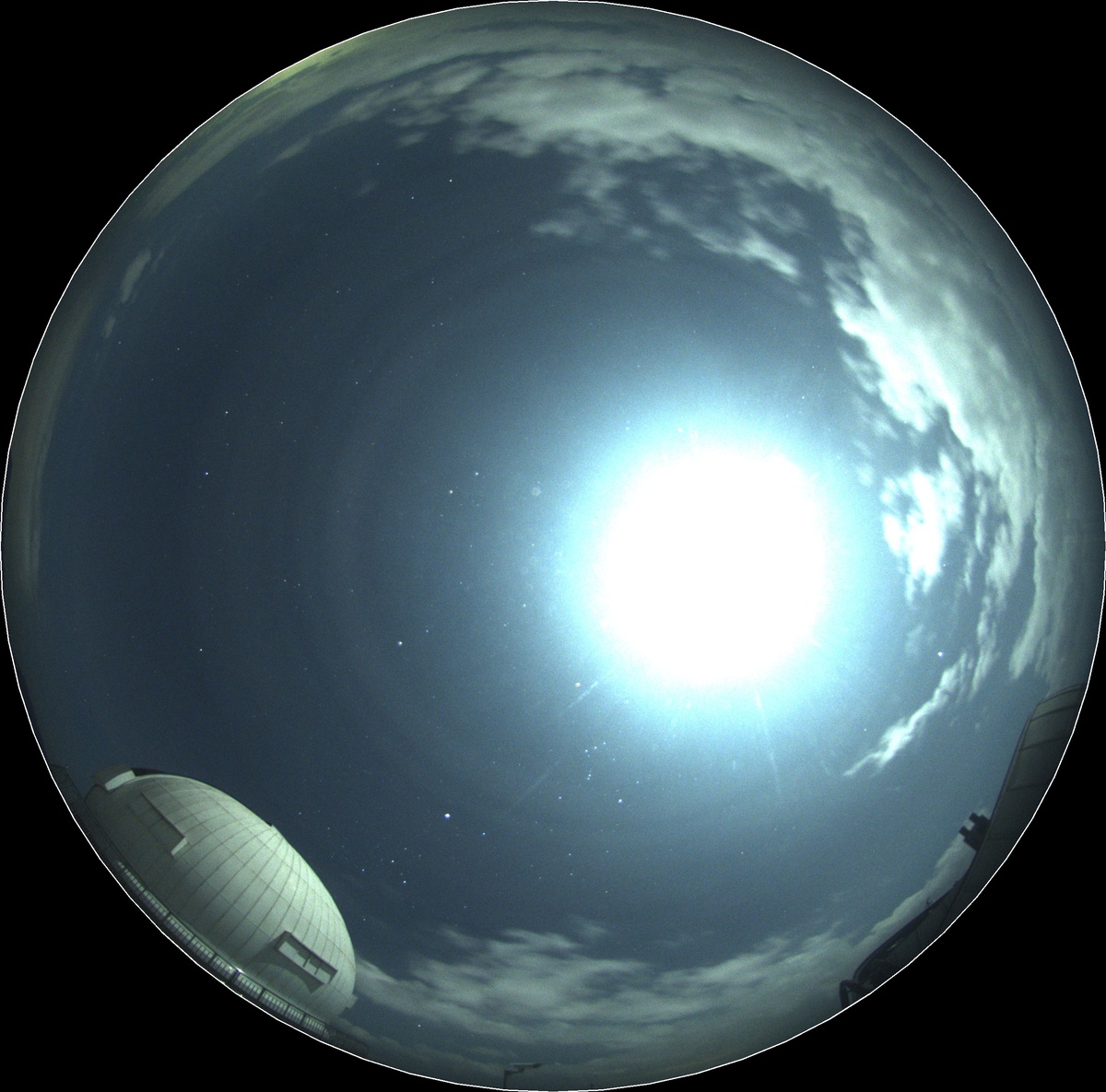} &  
        \includegraphics[width=1.8cm, height=1.8cm, keepaspectratio]{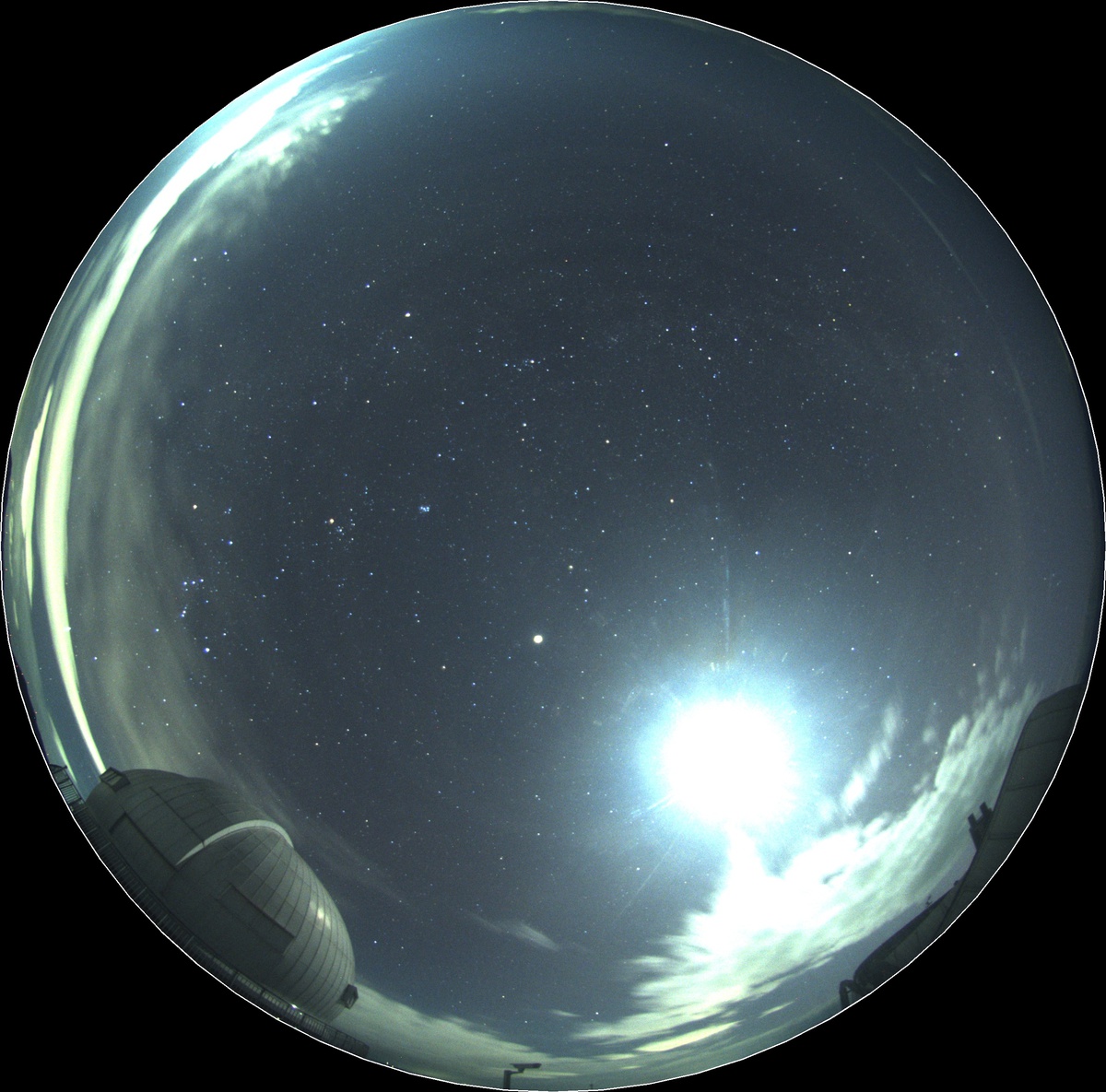} &  
        \includegraphics[width=1.8cm, height=1.8cm, keepaspectratio]{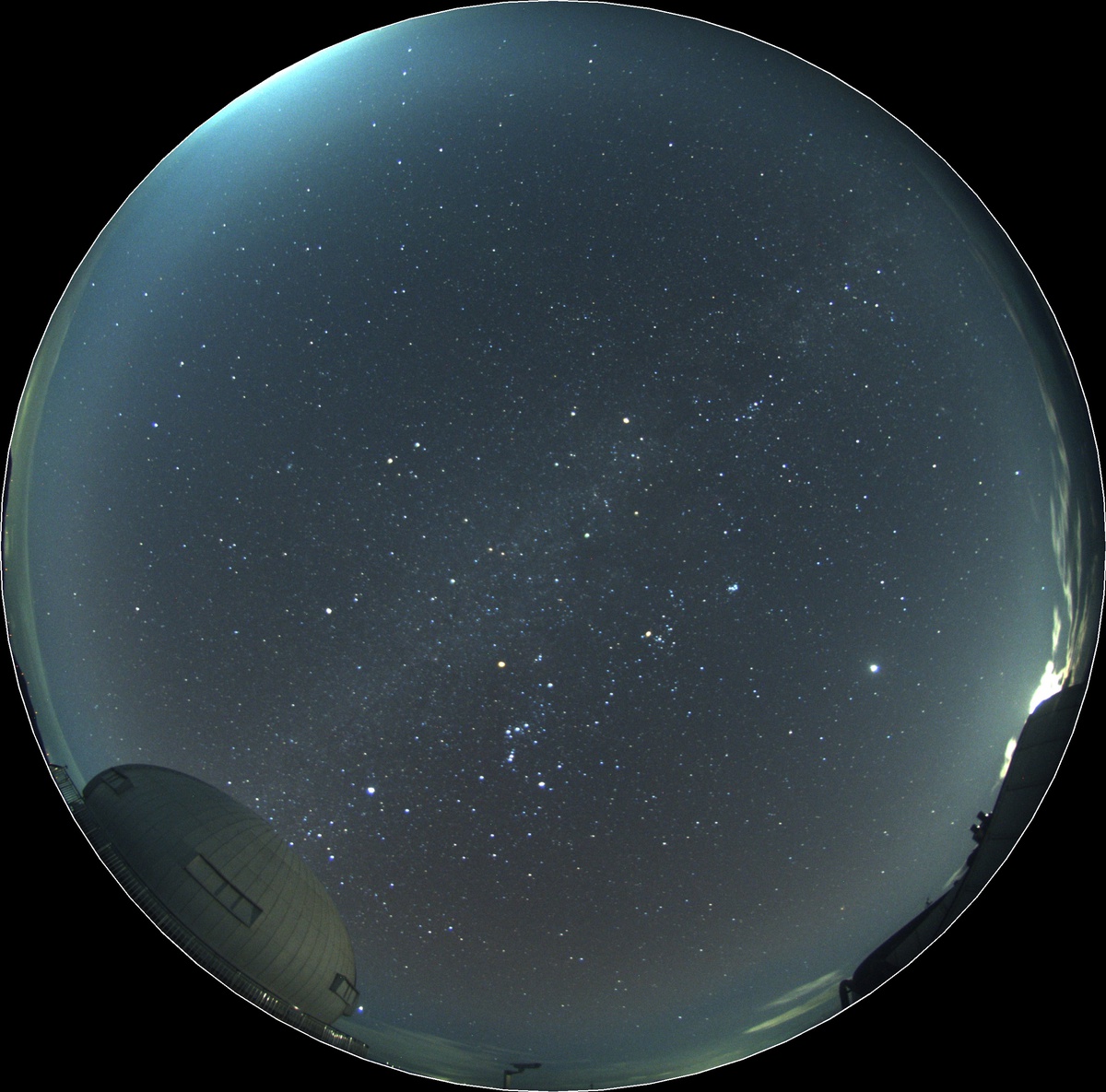} &  
        \includegraphics[width=1.8cm, height=1.8cm, keepaspectratio]{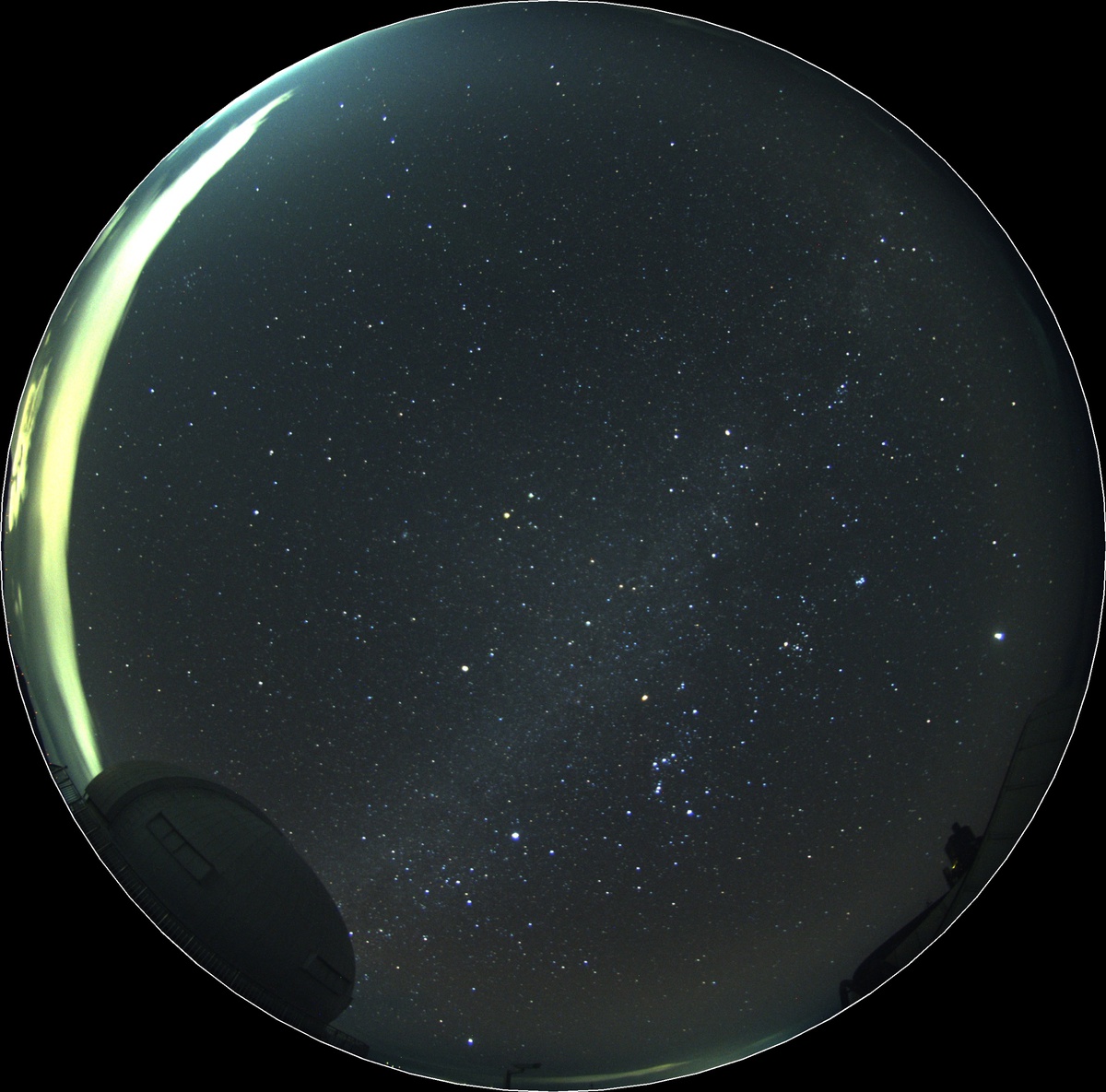} \\
        \midrule  
        \makecell[c]{\textbf{Mask}\\\textbf{Images}} &  
        \includegraphics[width=1.8cm, height=1.8cm, keepaspectratio]{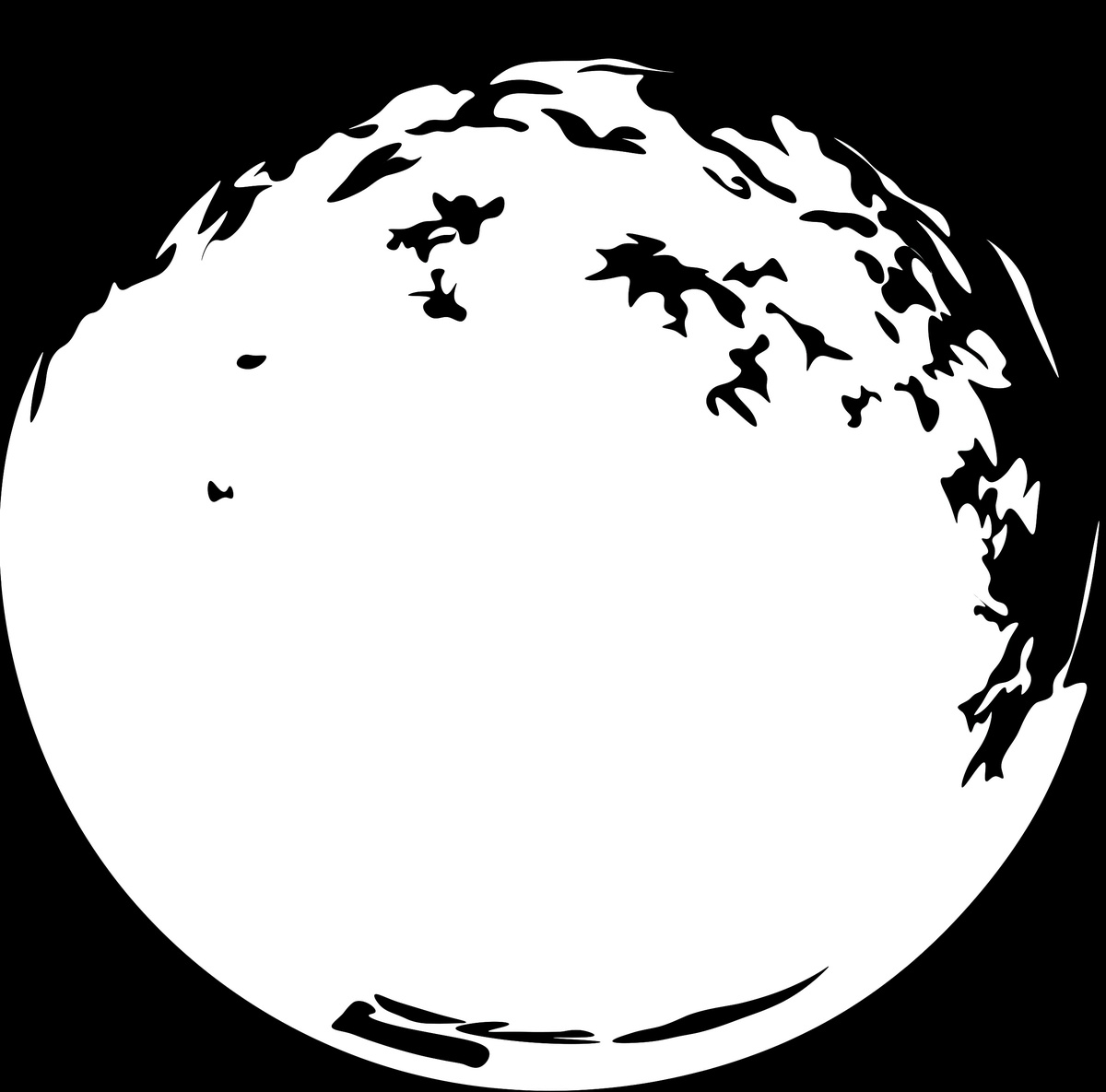} &  
        \includegraphics[width=1.8cm, height=1.8cm, keepaspectratio]{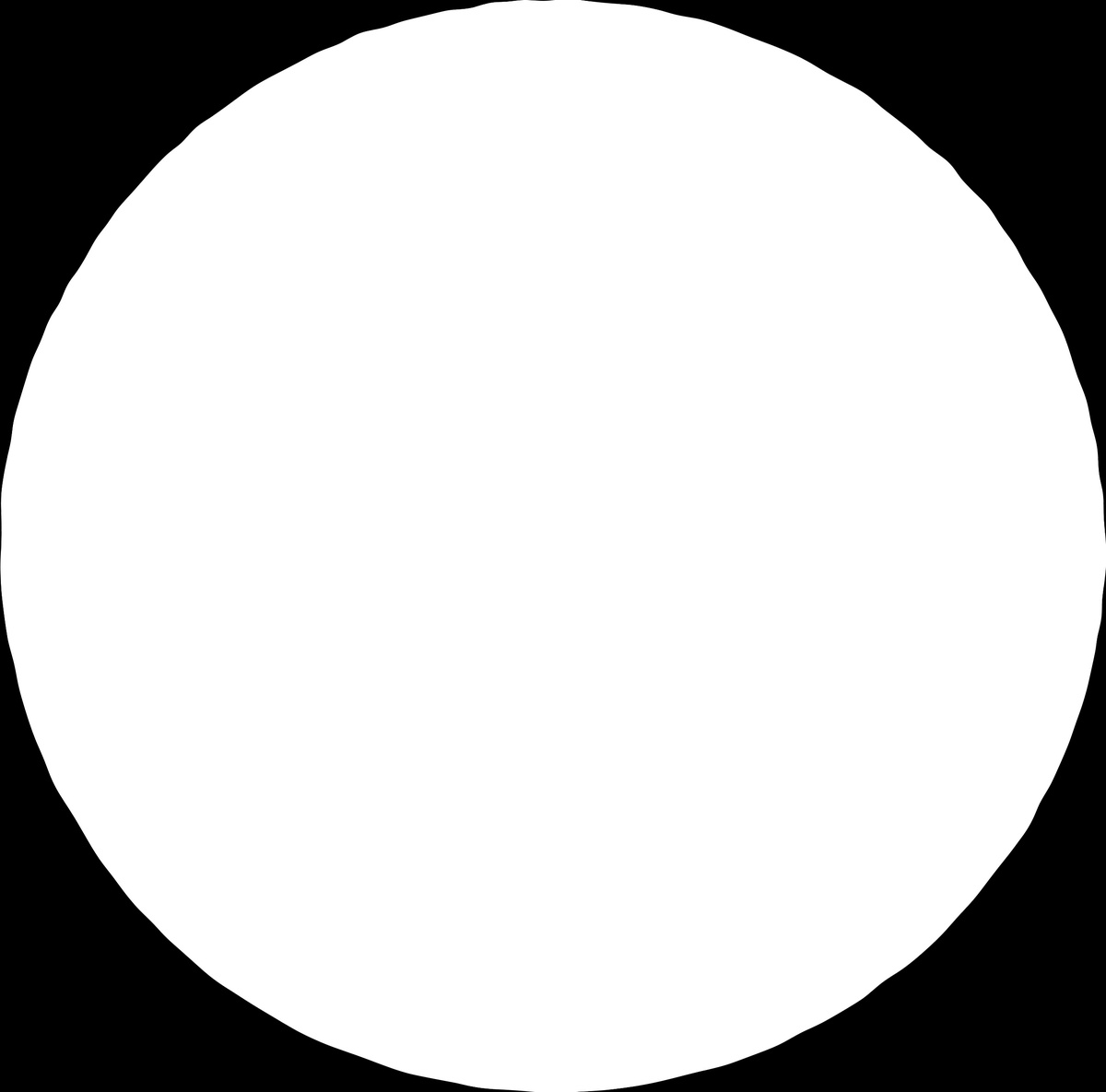} &  
        \includegraphics[width=1.8cm, height=1.8cm, keepaspectratio]{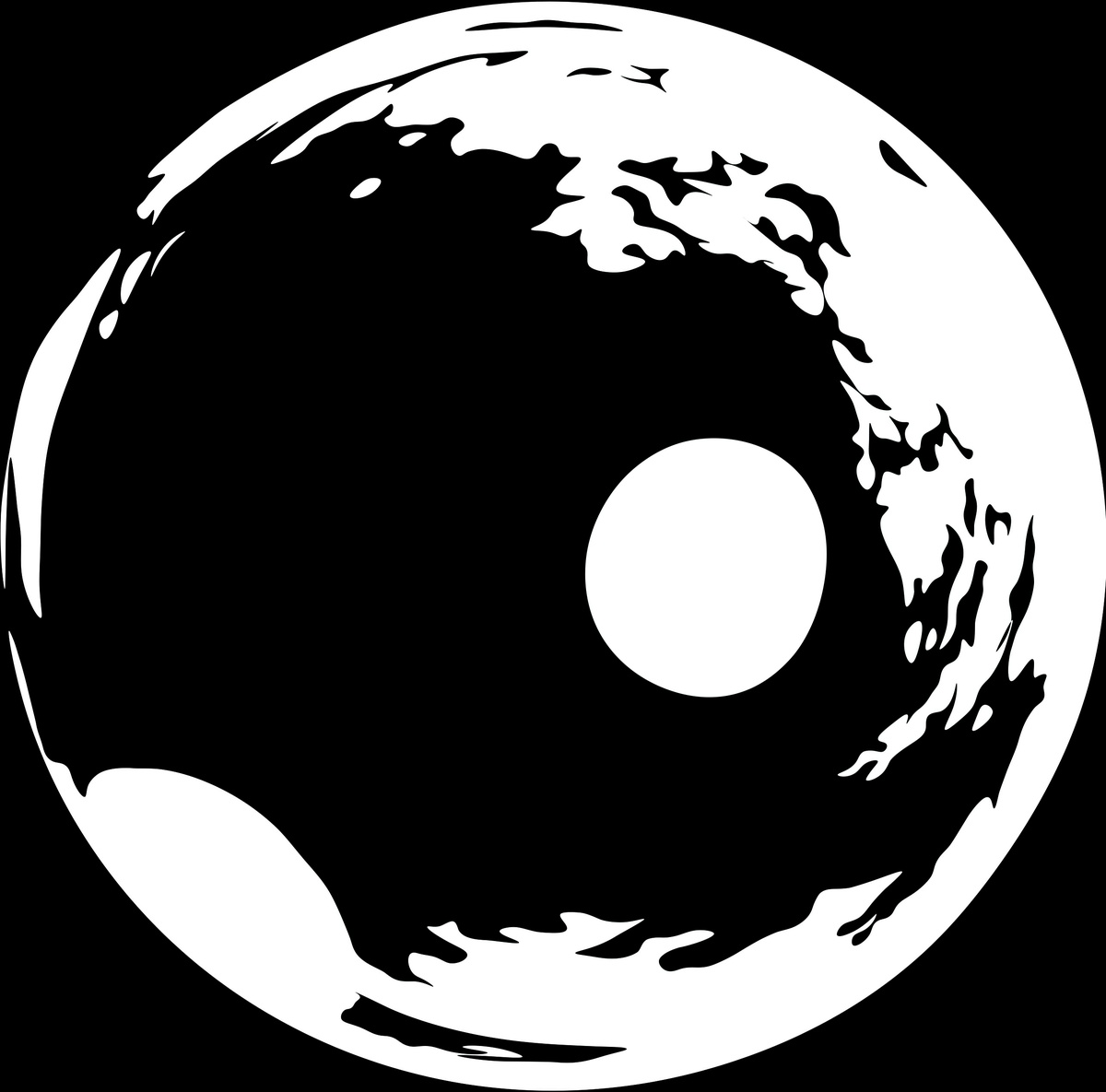} &  
        \includegraphics[width=1.8cm, height=1.8cm, keepaspectratio]{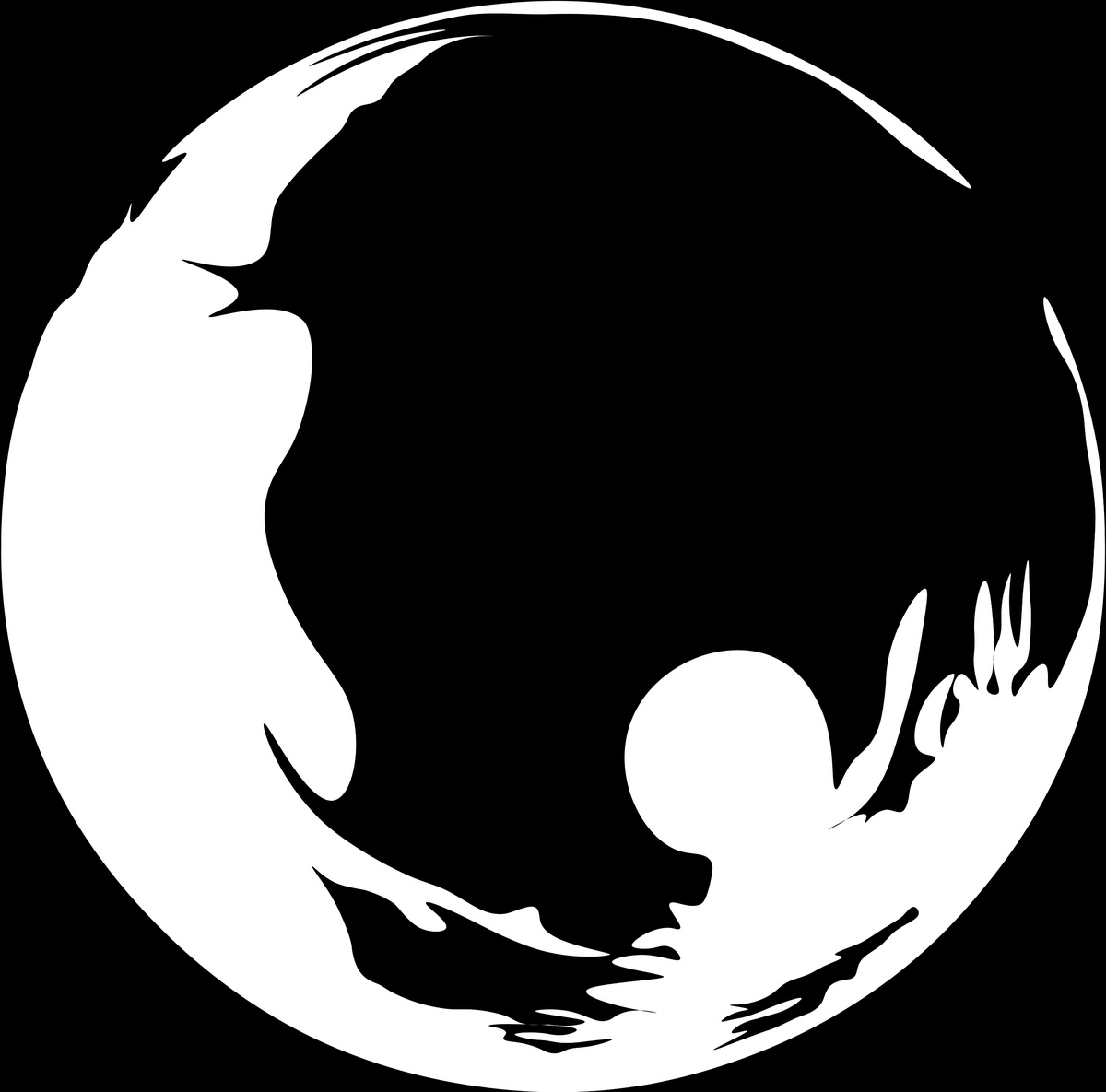} &  
        \includegraphics[width=1.8cm, height=1.8cm, keepaspectratio]{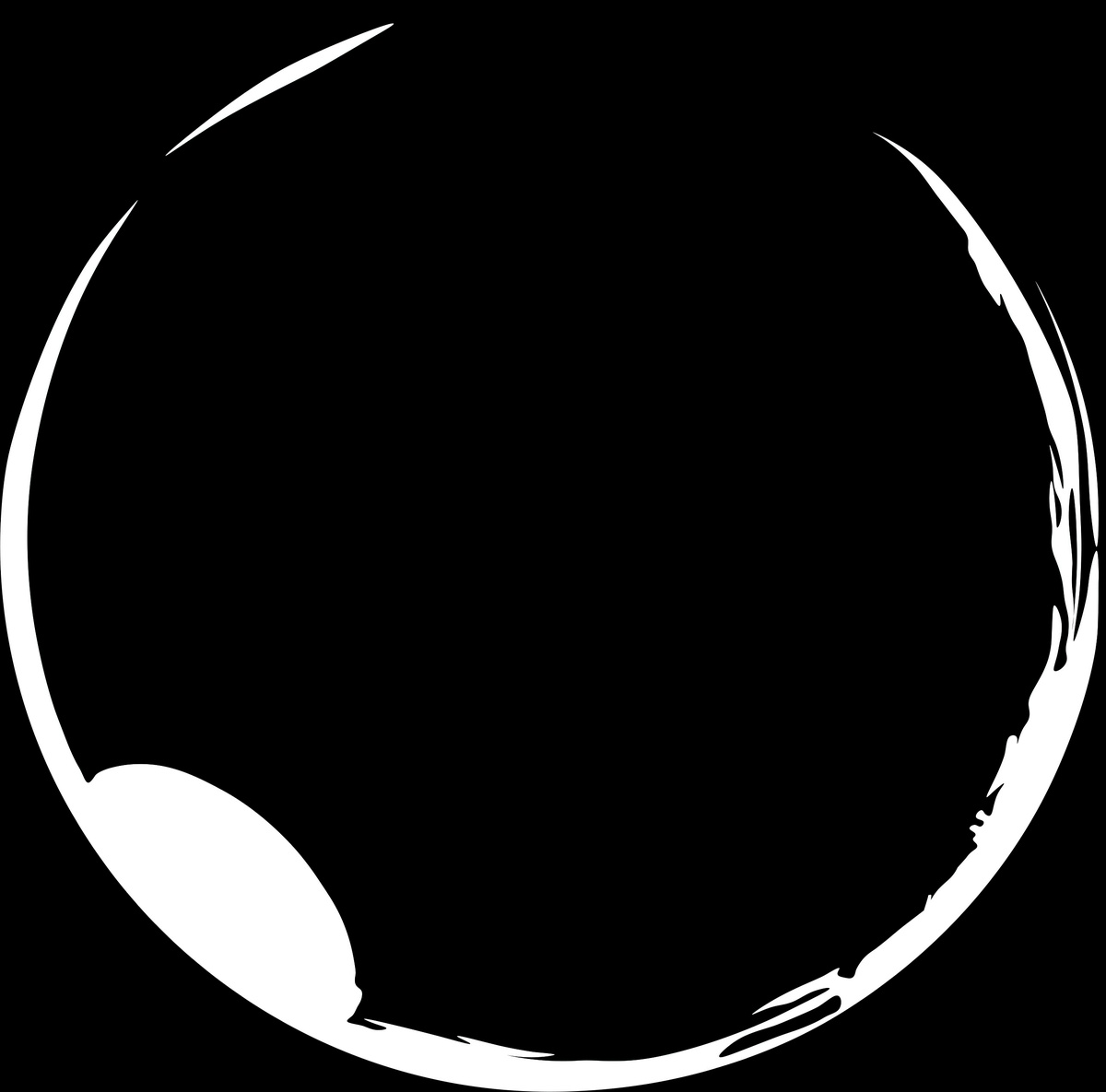} &  
        \includegraphics[width=1.8cm, height=1.8cm, keepaspectratio]{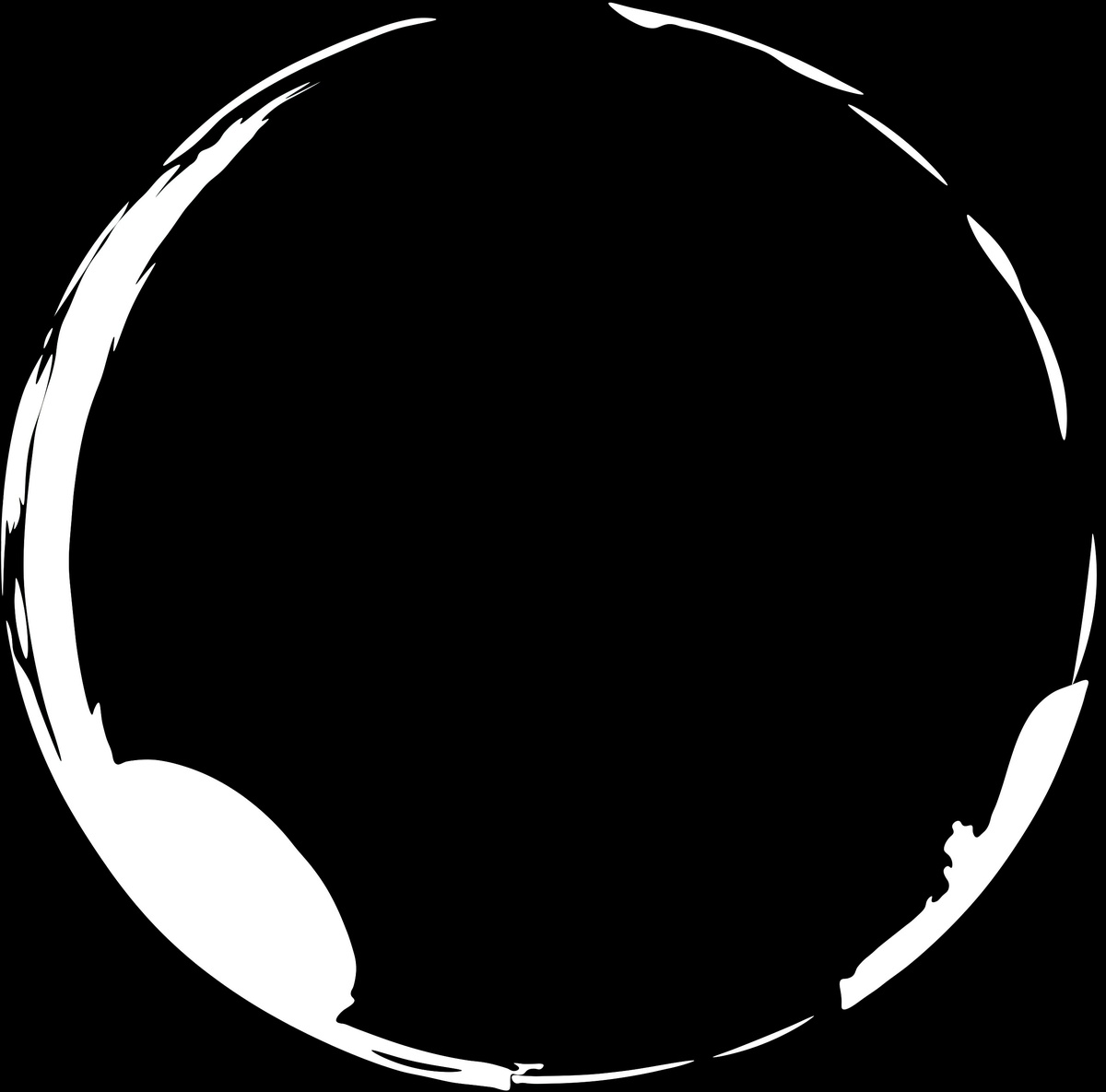} \\
        \midrule  
        \makecell[c]{\textbf{Mask}\\\textbf{Ratio}} &  
        \multicolumn{1}{c}{93.63\%} & \multicolumn{1}{c}{100\%} &  
        \multicolumn{1}{c}{22.15\%} & \multicolumn{1}{c}{21.47\%} &  
        \multicolumn{1}{c}{0.94\%} & \multicolumn{1}{c}{1.04\%} \\
        \bottomrule  
    \end{tabular}  
    \caption{  
        Examples of full sky image and mask image in three categories. Two representative samples are shown for each category. The mask ratio is given at the bottom for each sample.   
    }  
    \label{fig:Allsky_Weather_category}  
\end{figure*}

To provide intuitive insight into the classification scheme, Fig.~\ref{fig:Allsky_Weather_category} presents examples from each category. For each type, both the raw image and its corresponding mask are shown, along with the actual mask ratio (i.e., the proportion of unobservable regions). These examples visually demonstrate the diversity of sky conditions and the effectiveness of mask annotation in quantifying unobservable regions.  

\begin{table}  
\bc  
\begin{minipage}[]{100mm}  
\caption[]{Classification statistics of nighttime images captured by the all-sky camera based on the proportion of unobservable regions.\label{table:classification}}  
\end{minipage}  
\setlength{\tabcolsep}{4pt} 
\small   
\begin{tabular}{cccc}  
  \hline\noalign{\smallskip}  
  \textbf{Statistic} & \textbf{\makecell{High Unobservable\\Region}} & \textbf{\makecell{Moderate Unobservable\\Region}} & \textbf{\makecell{Low Unobservable\\Region}} \\
  \hline\noalign{\smallskip}   
\textbf{Images} & 724 & 596 & 680 \\
\textbf{Proportion} & 36.21\% & 29.77\% & 34.02\% \\ 
\textbf{\makecell{Criterion}} & $> 60\%$ & $2\%$--$60\%$ & $< 2\%$ \\
  \noalign{\smallskip}\hline  
\end{tabular}  
\ec  
\tablecomments{0.86\textwidth}{  
The statistics classify nighttime images based on the proportion of unobservable regions in each image. The columns correspond to the following categories: \textbf{High} (unobservable regions $>60\%$), \textbf{Moderate} ($2\%$--$60\%$), and \textbf{Low} ($<2\%$). Representative examples for each category are shown in Fig.~\ref{fig:Allsky_Weather_category}.  
}  
\end{table}

\begin{figure}[t]  
    \centering  
    \includegraphics[width=1.0\linewidth]{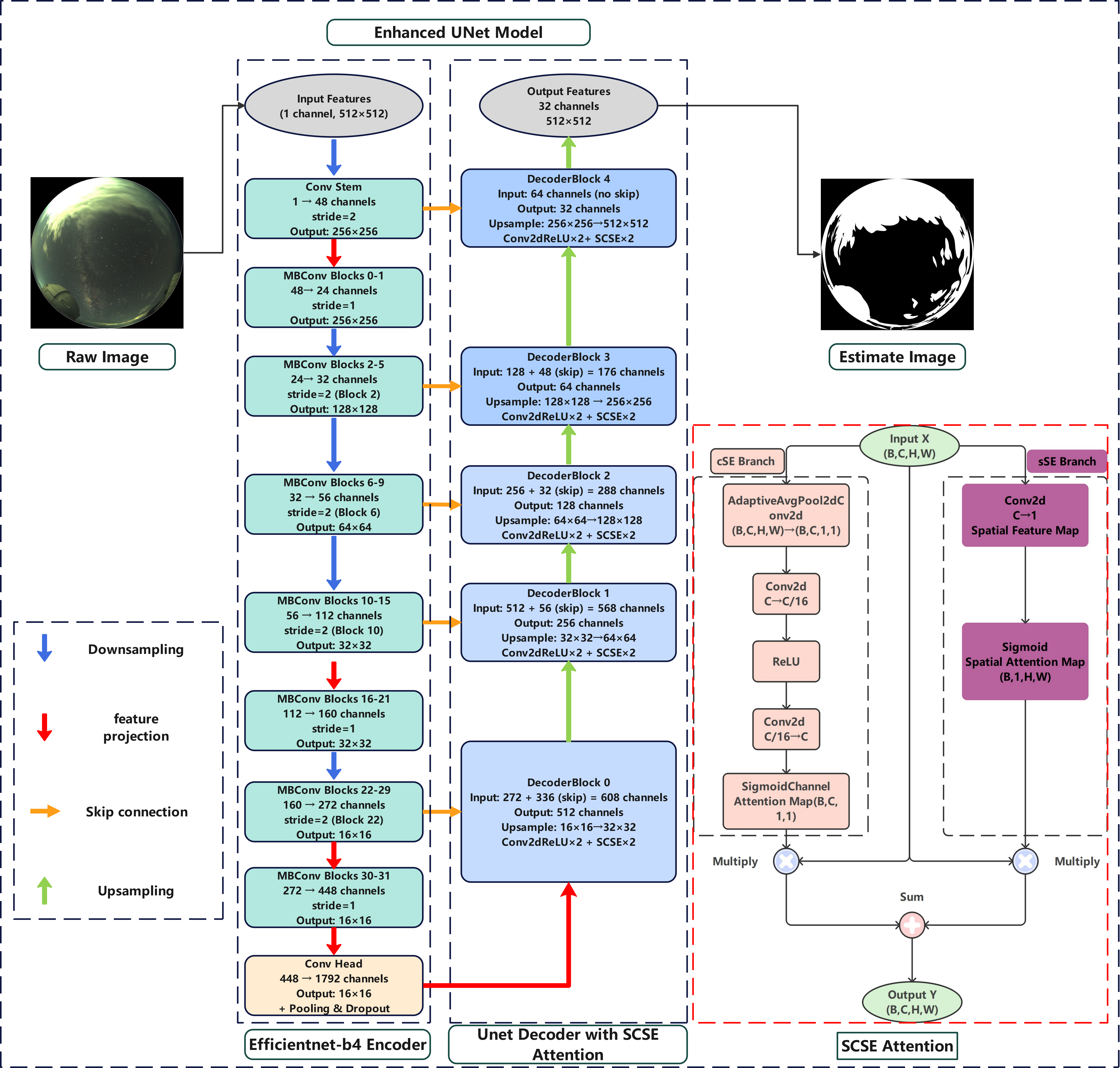}  
    \caption{The architecture of the proposed \textbf{Enhanced UNet}. The encoder is based on EfficientNet-B4, while the decoder incorporates SCSE attention modules for feature enhancement. Skip connections bridge the encoder and decoder to retain critical spatial details and contextual information. The output layer applies a Sigmoid activation for pixel-level precision.}  
    \label{fig:model_structure}  
\end{figure} 

\FloatBarrier 

\section{Unobservable Region Segmentation in Nighttime All-Sky Images} 
\label{sec:methodology}  

\subsection{Optimized Model Structure}  
This paper presents an improved model based on U-Net (\citealt{Ronneberger+etal+2015}), referred to as the \textbf{Enhanced UNet}. The architecture incorporates multiple enhancements over the baseline: an optimized encoder, an advanced decoder, effective regularization strategies, and a compound loss function to boost segmentation accuracy and model generalization.  

The encoder in the Enhanced UNet employs a pre-trained EfficientNet-B4 (\citealt{Tan+etal+2019}) as the feature extractor. By leveraging ImageNet-pretrained weights, the encoder can extract features more efficiently, improve classification accuracy, and significantly reduce training time, ensuring the model learns diverse and meaningful representations from the input data.

In the decoder, the model incorporates SCSE attention modules (\citealt{Hu+etal+2017}), which dynamically recalibrate channel and spatial feature weights, enabling the network to focus more on salient information. A custom attention block can be incorporated if finer multi-scale feature extraction is needed. To further boost generalization and prevent overfitting, a Dropout layer is applied after the decoder’s final output. Skip connections throughout the network ensure both global and local contextual information are preserved, which improves pixel-level segmentation precision. The output layer utilizes a single-channel Sigmoid activation to generate pixel-wise probability maps.  

To address the specific challenges of segmenting unobservable sky regions, the Enhanced UNet is trained using a \textbf{combined loss function} (\citealt{Jadon+2020}), which is a weighted sum of Binary Cross-Entropy (BCE) loss, Dice loss, and Intersection over Union (IoU) loss. This approach leverages the strengths of each loss---BCE for pixel-wise classification, Dice for class imbalance, and IoU for region-level overlap---to optimize different facets of the segmentation task. The combined loss is formulated as: \[   \text{CombinedLoss} = w_1\cdot\text{loss}_{\text{BCE}} + w_2\cdot\text{loss}_{\text{Dice}} + w_3\cdot\text{loss}_{\text{IoU}},   \] where \(w_1\), \(w_2\) and \(w_3\) are tunable hyperparameters. 

The overall architecture, as shown in Fig.~\ref{fig:model_structure}, is tailored to maximize segmentation accuracy and robust feature representation, especially under complex nighttime sky conditions.

\subsection{Experimental Configuration}  

The experimental setup employed in this study is detailed in Table~\ref{tab:experiment-configuration}. The Enhanced UNet model is trained and evaluated on a dataset comprising 2,000 high-resolution nighttime all-sky images. Each image is an RGB color image with a resolution of $2568 \times 2538$ pixels. Ground truth segmentation masks are manually annotated using image editing software, resulting in binary (black and white) masks precisely aligned with the original images. For model training, all input images are converted to grayscale, while the corresponding binary masks served as segmentation labels. The dataset is randomly divided into a training set (90\%, 1,800 images) and a test set (10\%, 200 images), ensuring both robust model learning and unbiased performance evaluation.  

The network architecture is based on a pretrained EfficientNet-B4 encoder and a U-Net decoder enhanced with spatial and channel squeeze-and-excitation (SCSE) attention mechanisms. Model optimization is performed using the RAdam optimizer, with an initial learning rate of $2 \times 10^{-4}$ and a weight decay of $1 \times 10^{-5}$. A cosine annealing warm restarts scheduler ($T_0 = 10$, $T_\text{mult} = 2$, minimum learning rate $1 \times 10^{-6}$) is employed to dynamically adjust the learning rate throughout training. The model is trained for up to 150 epochs with a batch size of 16, and early stopping is triggered if validation performance does not improve for 20 consecutive epochs. This experimental setup provides a robust and reproducible basis for subsequent model training and evaluation. 

To objectively select the optimal loss weights, we performed a systematic grid search across the weight vector 
$(w_1, w_2, w_3)$ with a step size of $0.1$, enforcing the constraint $w_1 + w_2 + w_3 = 1$, $w_i \geq 0$. 
Each configuration was trained with identical data splits, optimization hyper\-parameters, and data augmentations, 
ensuring a fair and unbiased comparison. The validation set IoU served as the primary selection metric, 
with F1 score as a tie\-breaker. Early stopping based on validation IoU (patience = $20$ epochs) was applied to 
prevent over\-fitting. Each setting was evaluated over three random initializations, and we report the mean and 
standard deviation for all metrics (see Table~\ref{table:results} and Fig.~\ref{fig:performance_evaluation}). 
To verify that the optimum is not due to grid discretization, a finer search (step = $0.05$) was conducted around 
the best weight, confirming ranking stability. The hold-out test set, which was never used during training or model 
selection, was used for final model assessment.

\begin{table}[t]  
\bc  
\begin{minipage}[]{100mm}  
\caption[]{Experimental Environment \label{tab:experiment-configuration}}  
\end{minipage}  
\setlength{\tabcolsep}{2pt}  
\small  
\begin{tabular}{cc}  
\noalign{\smallskip}\hline\noalign{\smallskip}  
\textbf{Device} & \textbf{Configuration} \\
\noalign{\smallskip}\hline\noalign{\smallskip}  
Operating System & Windows 11 Chinese Home Edition \\
Processor & Intel(R) Core(TM) i9-14900K (3.20 GHz, 24 cores, 32 threads) \\
RAM & 64 GB \\
PyTorch Version & 2.5.1 \\
GPU & NVIDIA GeForce RTX 4090 D \\
CUDA Version & 11.8 \\
\noalign{\smallskip}\hline  
\end{tabular}  
\ec  
\end{table}  

\subsection{Experimental Results}  

\begin{table}[t]  
\begin{center}  
\caption[]{Comparison of the U-Net baseline and Enhanced U-Net on the expanded dataset. Loss parameters ($w_1$, $w_2$, $w_3$) and evaluation metrics (IoU, Precision, Recall, F1 Score) are shown. All Enhanced U-Net settings outperform the baseline; the best values are highlighted in bold. \label{table:results}}  
\begin{tabular}{lccccccc}  
  \hline\noalign{\smallskip}  
\textbf{Model}              & \textbf{\(w_1\)} & \textbf{\(w_2\)} & \textbf{\(w_3\)} & \textbf{IoU}   & \textbf{Precision} & \textbf{Recall} & \textbf{F1 Score} \\
  \hline\noalign{\smallskip}  
U-Net Baseline              & -               & -               & -               & 0.9025         & 0.9401             & 0.9441          & 0.9320 \\
\hline  
\multirow{9}{*}{Enhanced U-Net}  
                            & 0.1             & 0.3             & 0.6             & 0.9164         & 0.9531             & 0.9572          & 0.9468 \\
                            & 0.3             & 0.3             & 0.4             & 0.9178         & 0.9544             & 0.9598          & 0.9506 \\
                            & 0.5             & 0.3             & 0.2             & 0.9162         & 0.9527             & 0.9575          & 0.9465 \\
                            & 0.2             & 0.5             & 0.3             & 0.9159         & 0.9523             & 0.9569          & 0.9460 \\
                            & 0.4             & 0.4             & 0.2             & 0.9160         & 0.9525             & 0.9571          & 0.9462 \\
                            & \textbf{0.4}    & \textbf{0.2}    & \textbf{0.4}    & \textbf{0.9212} & \textbf{0.9564}    & \textbf{0.9602} & \textbf{0.9537} \\
                            & 1.0             & 0.0             & 0.0             & 0.9141         & 0.9509             & 0.9555          & 0.9443 \\
                            & 0.0             & 1.0             & 0.0             & 0.9146         & 0.9512             & 0.9558          & 0.9450 \\
                            & 0.0             & 0.0             & 1.0             & 0.9151         & 0.9519             & 0.9562          & 0.9457 \\
  \noalign{\smallskip}\hline  
\end{tabular}  
\end{center}  
\end{table}

\begin{figure*}[t]  
    \centering    
    \includegraphics[width=1.0\textwidth]{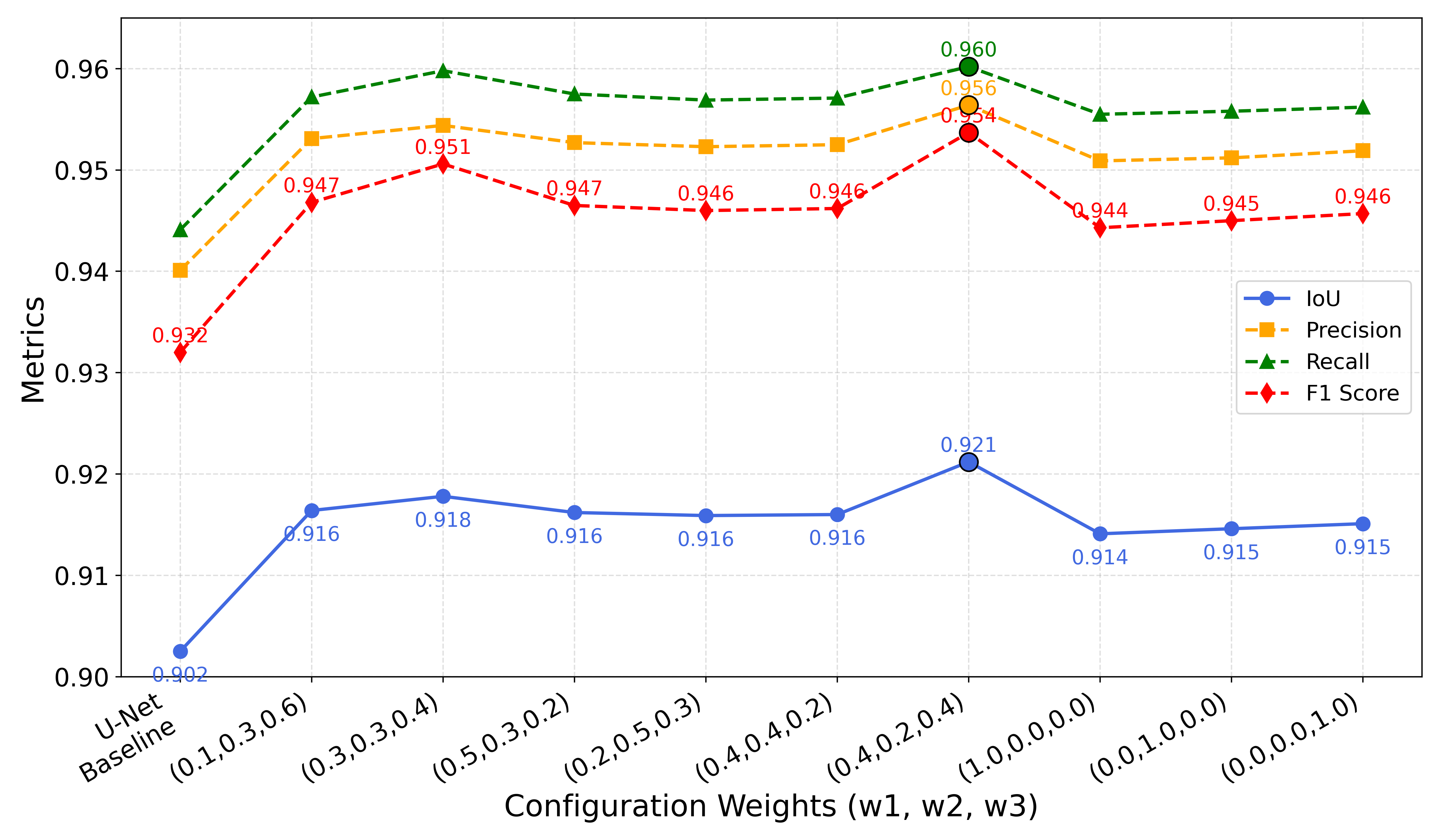}  
    \caption{  
        Performance evaluation of the U-Net baseline and Enhanced UNet configurations on the expanded dataset.  
        The Enhanced UNet achieves consistently better IoU, Precision, Recall, and F1 Score.  
        The configuration (\(w_1 = 0.4, w_2 = 0.2, w_3 = 0.4\)) yields the best results.  
    }  
    \label{fig:performance_evaluation}  
\end{figure*} 

\begin{figure*}[t]  
    \centering   
    \setlength{\abovecaptionskip}{4pt}   
    \setlength{\tabcolsep}{2pt}  
    \renewcommand{\arraystretch}{1.06}  
    \small  
    \begin{tabular}{%
        >{\centering\arraybackslash}m{2.3cm}  
        >{\centering\arraybackslash}m{2.8cm}  
        >{\centering\arraybackslash}m{2.8cm}  
        >{\centering\arraybackslash}m{2.8cm}  
    }  
        \toprule  
        \textbf{Category} &   
        \textbf{High Unobservable Region} &   
        \textbf{Moderate Unobservable Region} &   
        \textbf{Low Unobservable Region} \\
        \midrule  
        \textbf{Raw Image} &  
        \includegraphics[width=0.9\linewidth]{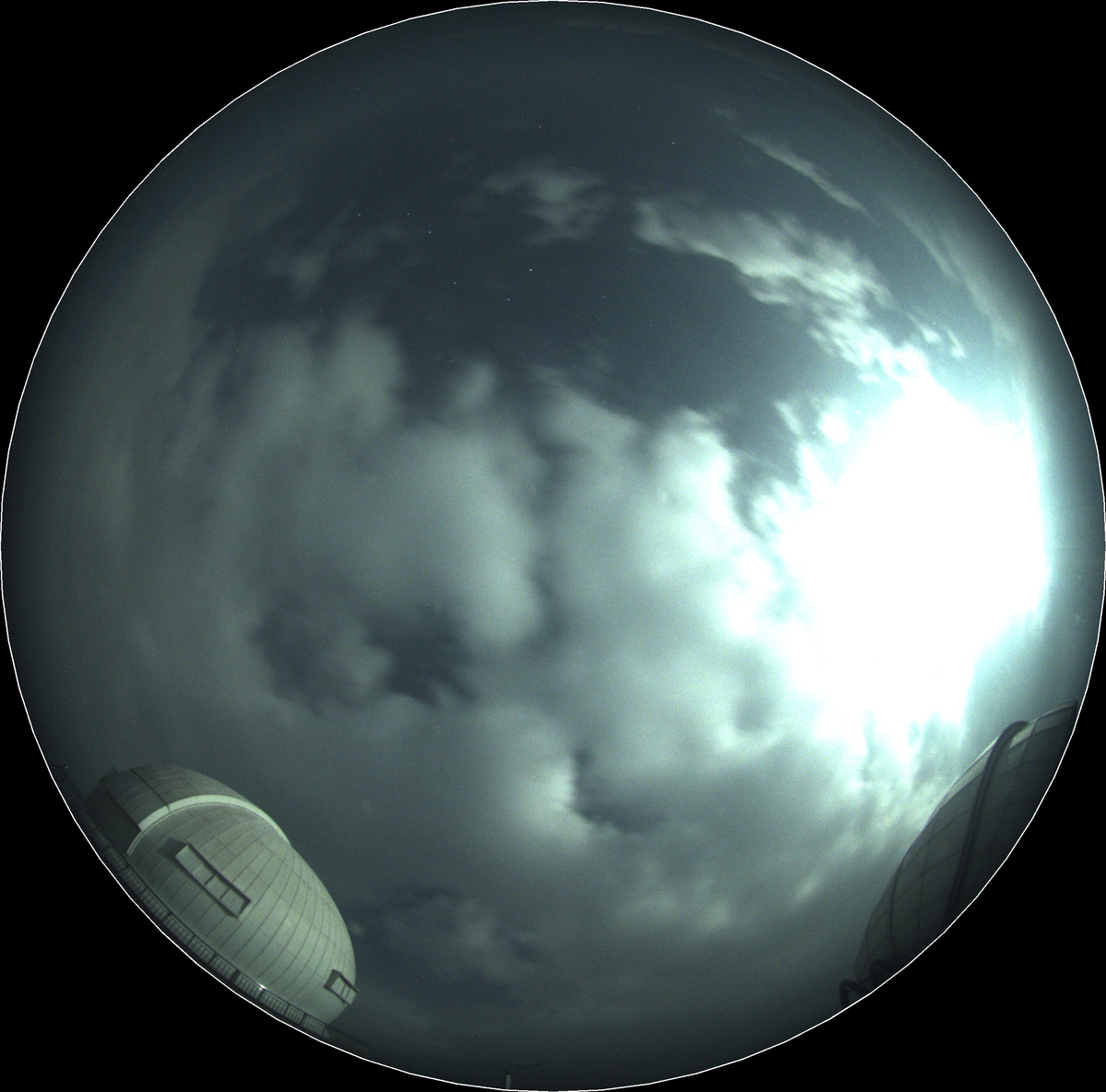} &  
        \includegraphics[width=0.9\linewidth]{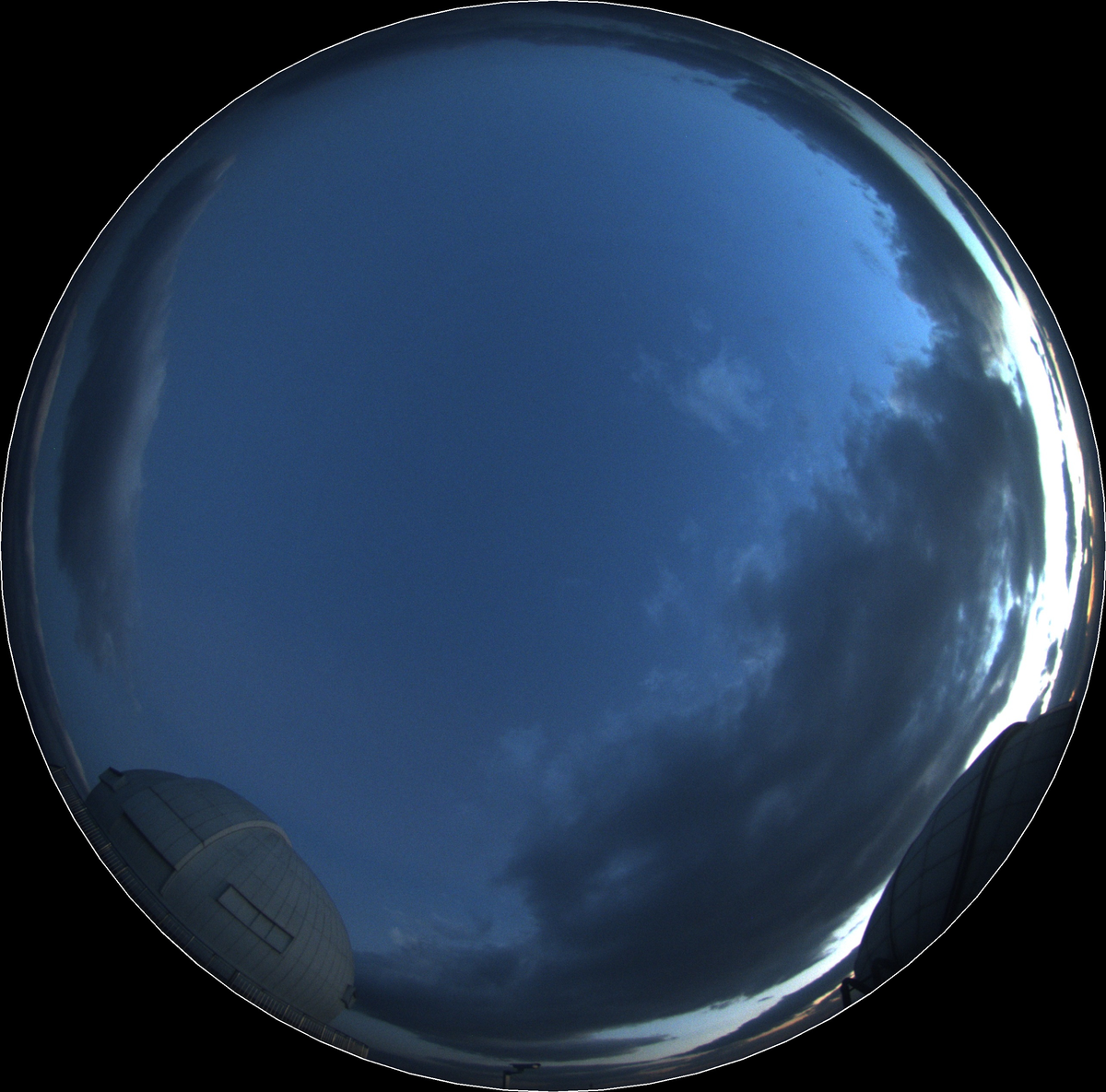} &  
        \includegraphics[width=0.9\linewidth]{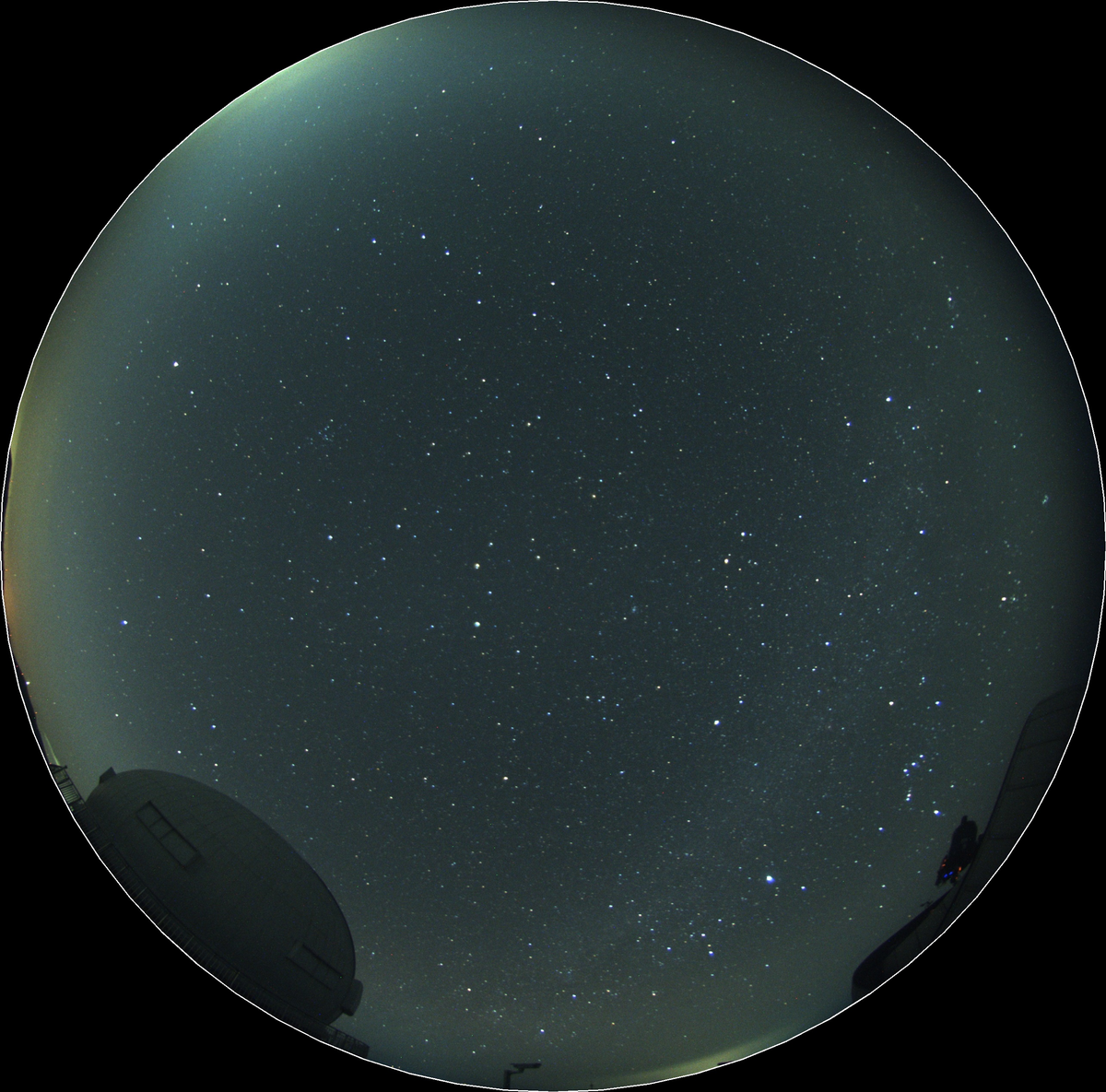} \\
        \midrule  
        \textbf{Model Prediction} &  
        \includegraphics[width=0.9\linewidth]{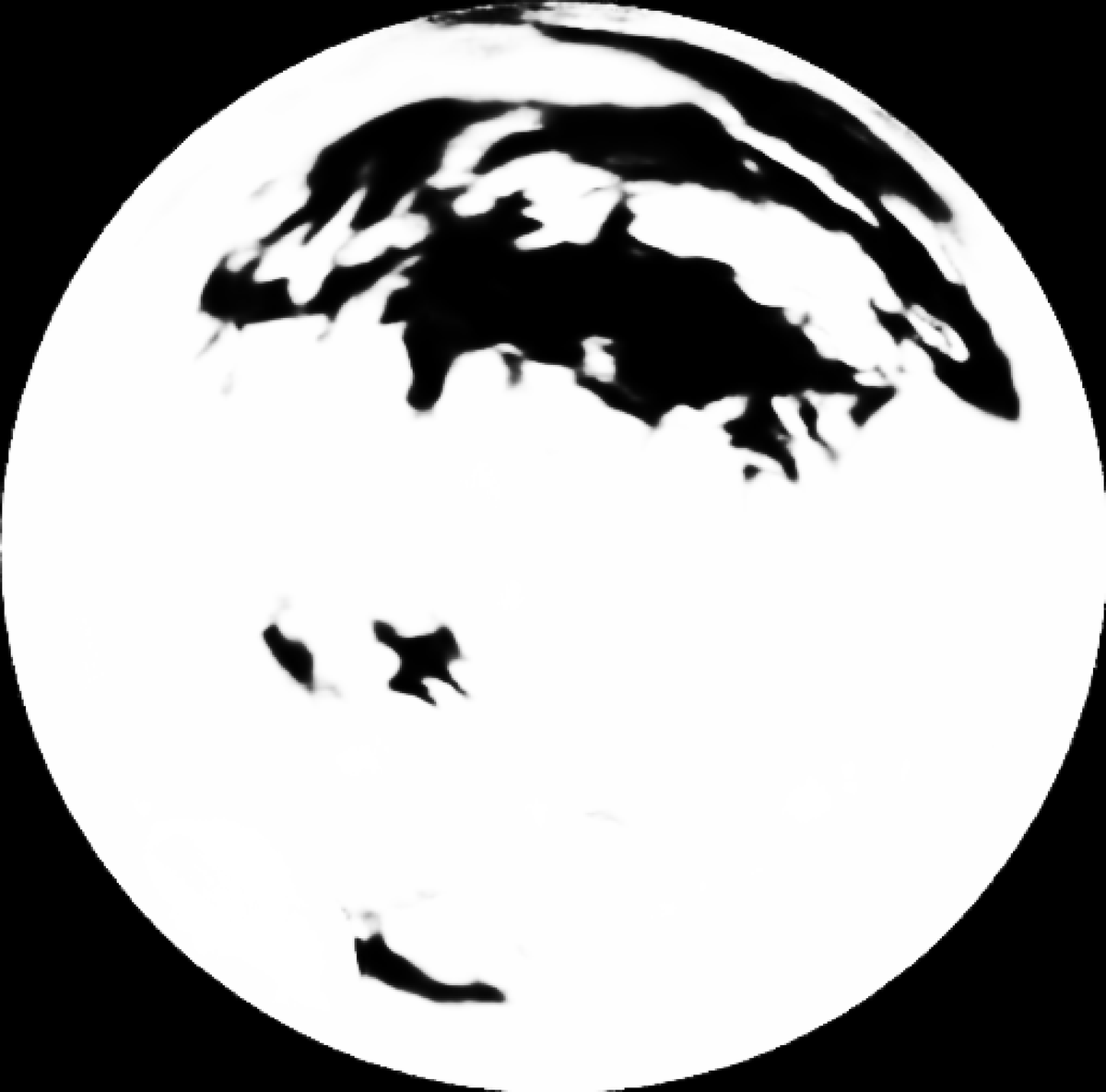} &  
        \includegraphics[width=0.9\linewidth]{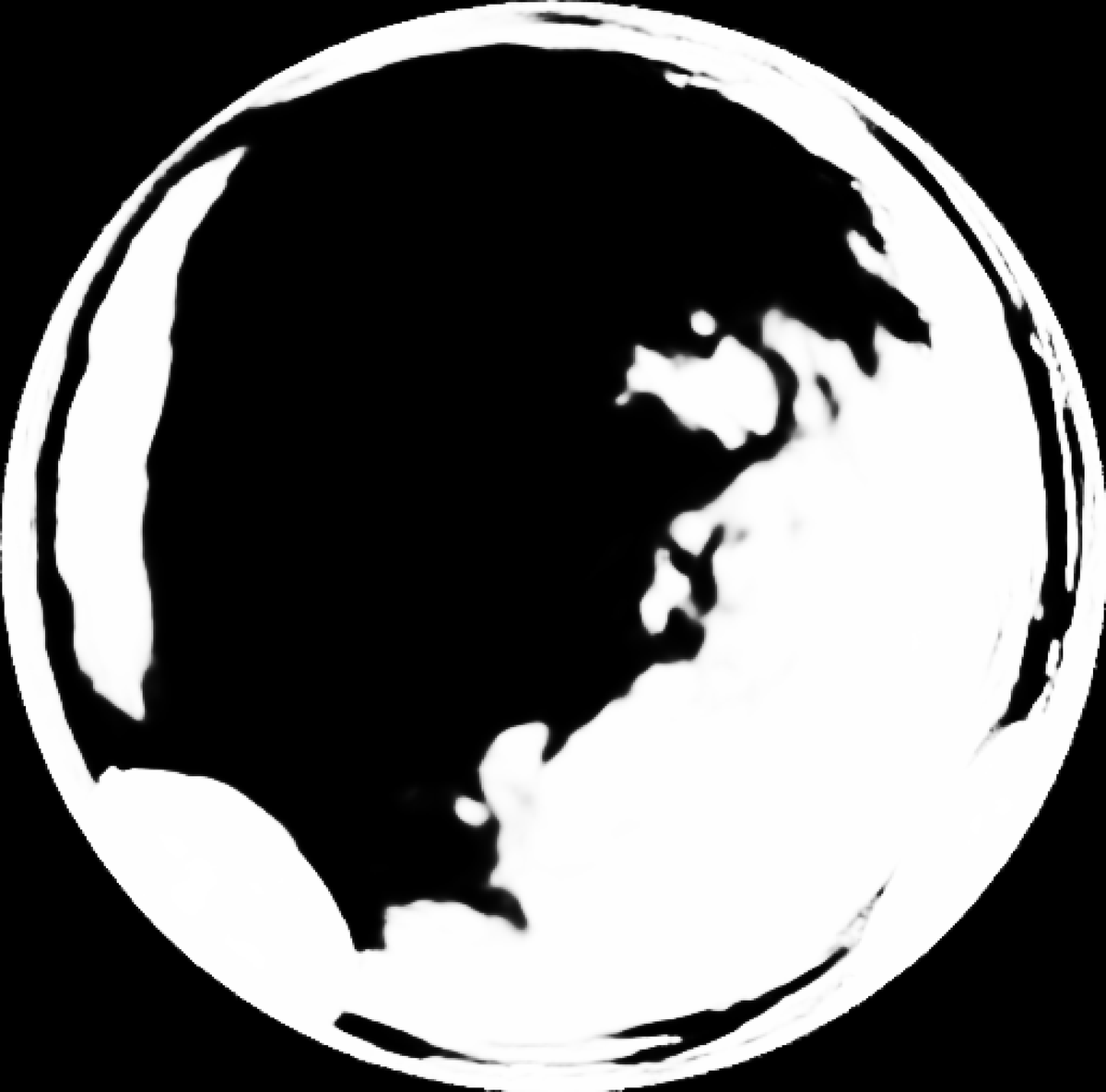} &  
        \includegraphics[width=0.9\linewidth]{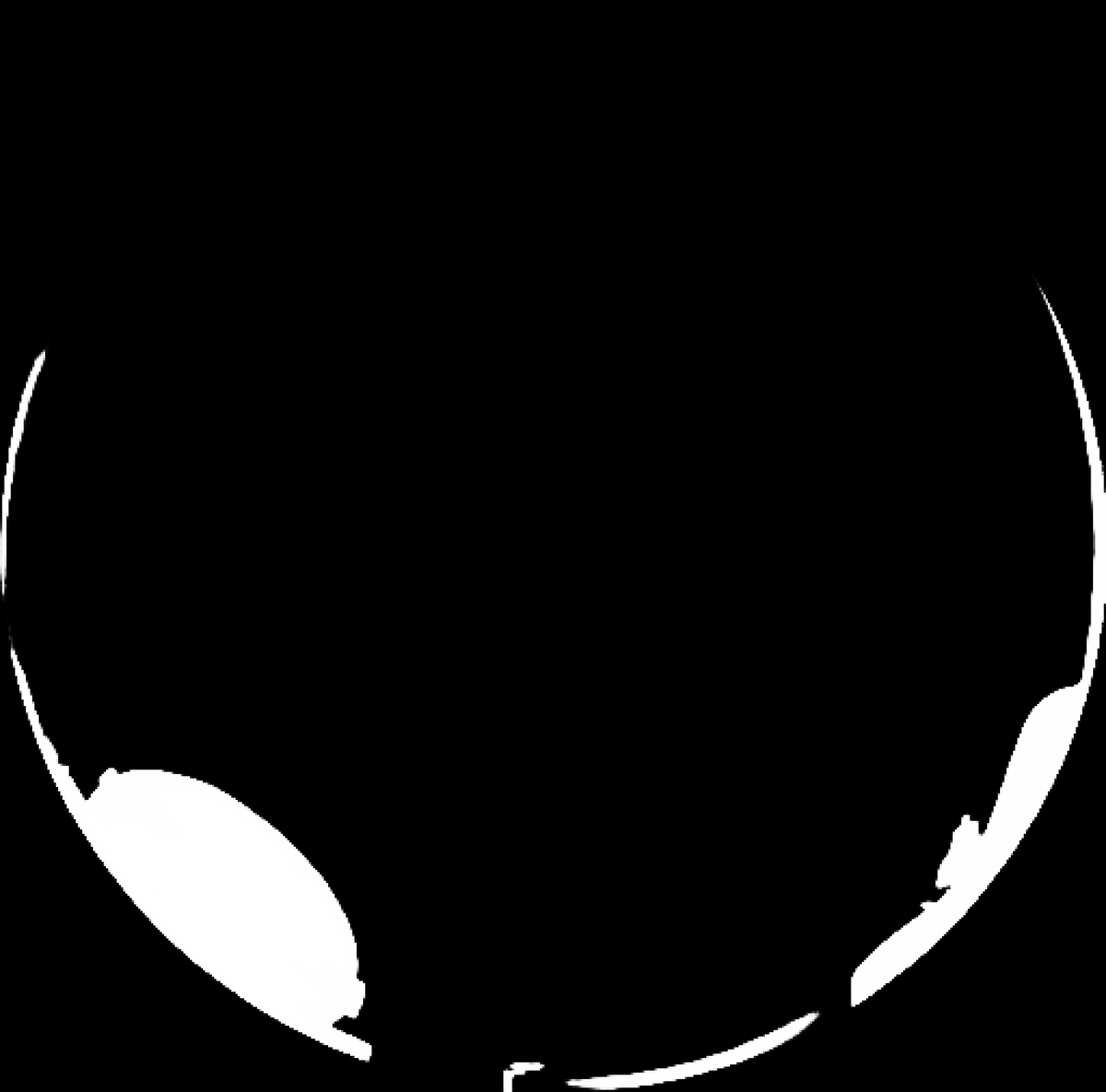} \\
        \midrule  
        \textbf{Ground Truth} &  
        \includegraphics[width=0.9\linewidth]{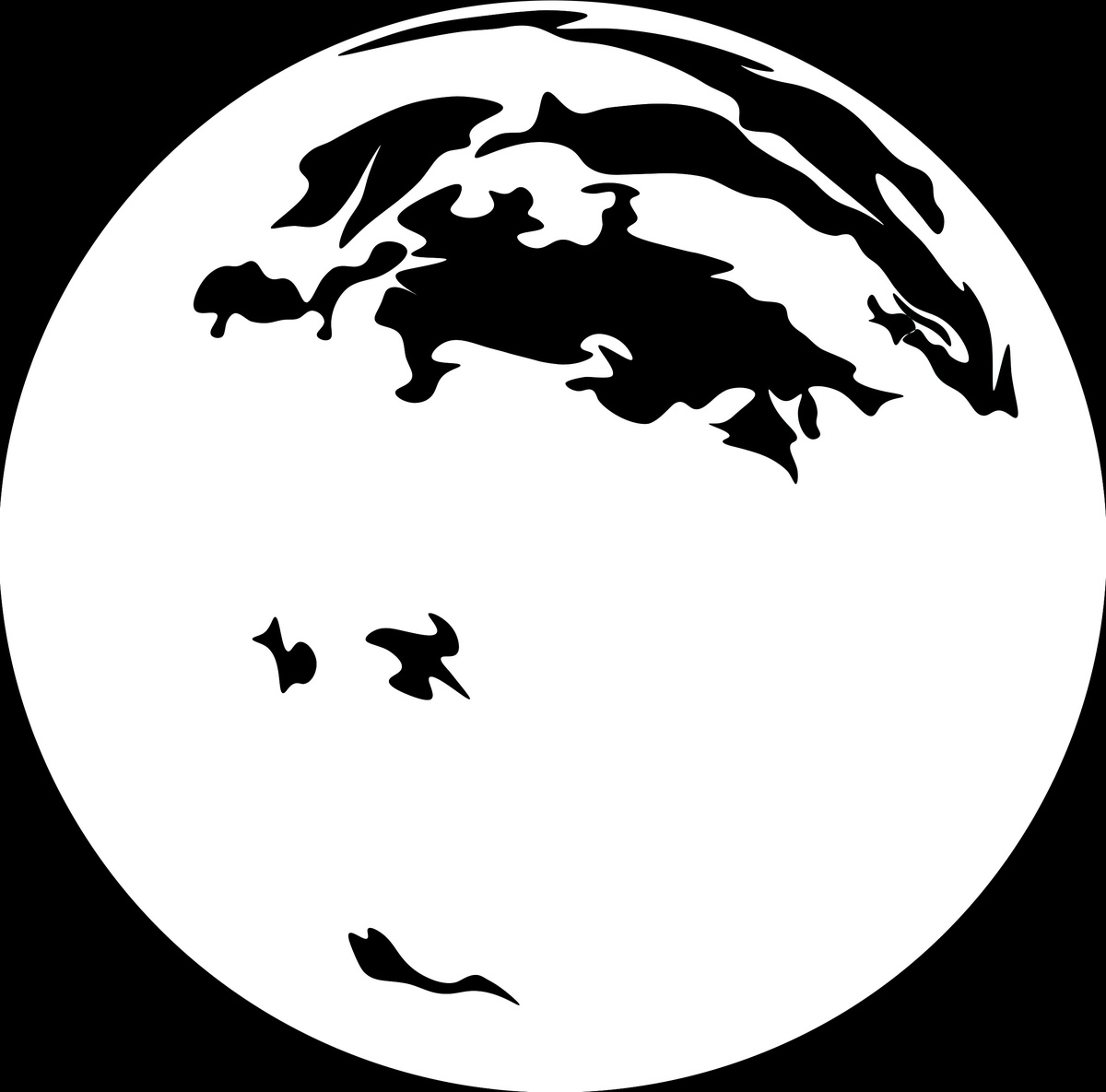} &  
        \includegraphics[width=0.9\linewidth]{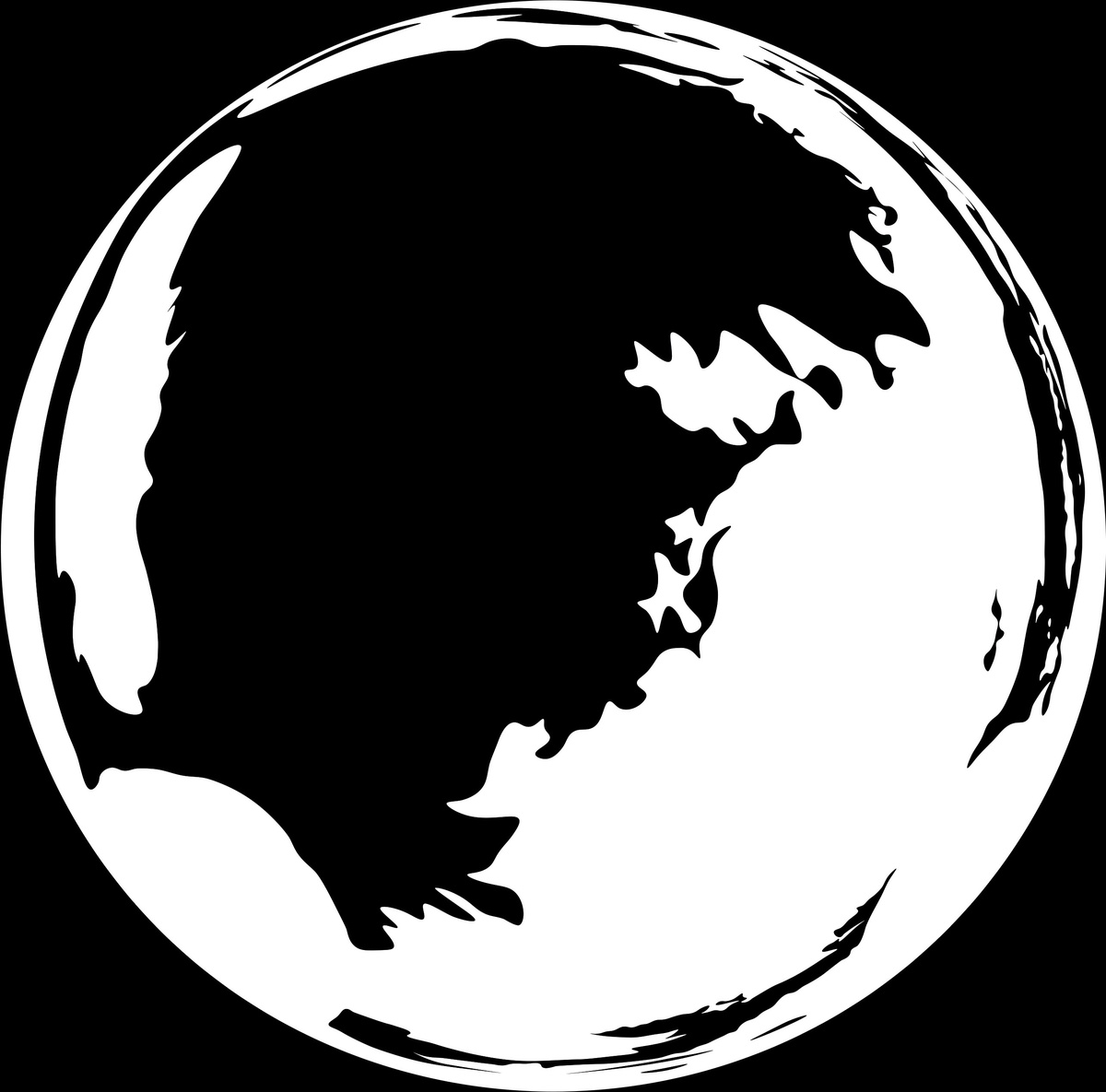} &  
        \includegraphics[width=0.9\linewidth]{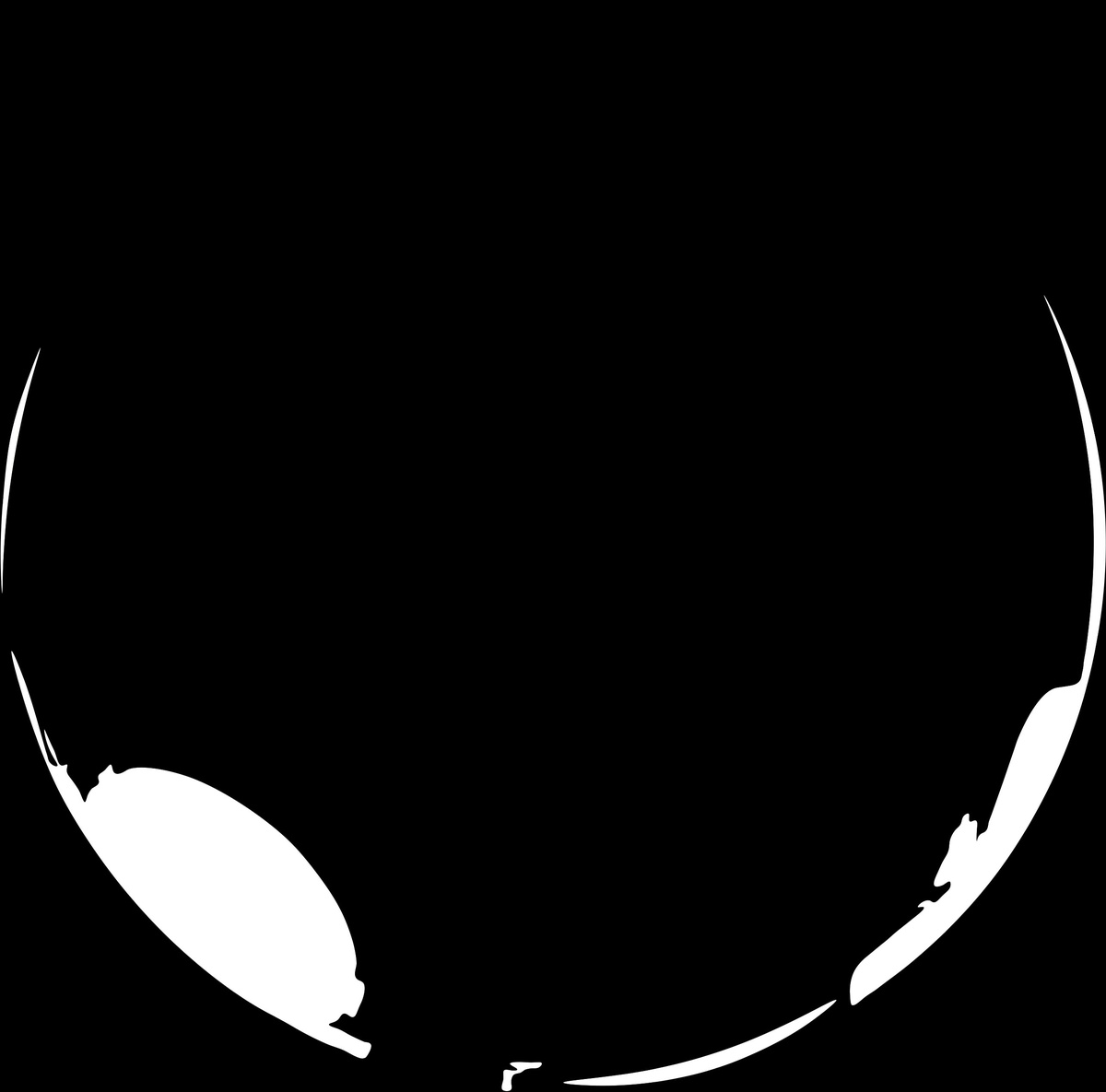} \\
        \midrule  
        \textbf{Overlay Image} &  
        \includegraphics[width=0.9\linewidth]{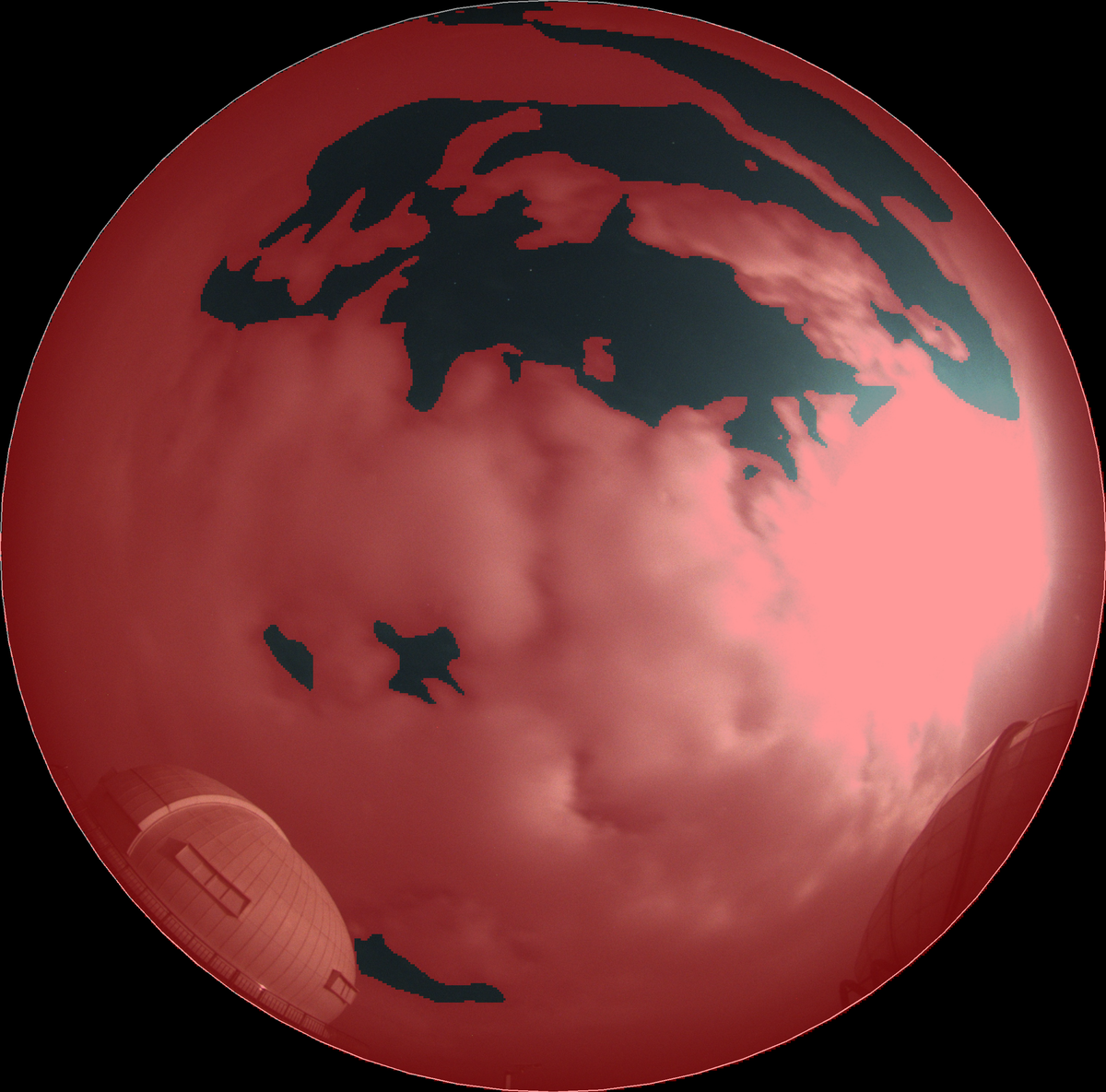} &  
        \includegraphics[width=0.9\linewidth]{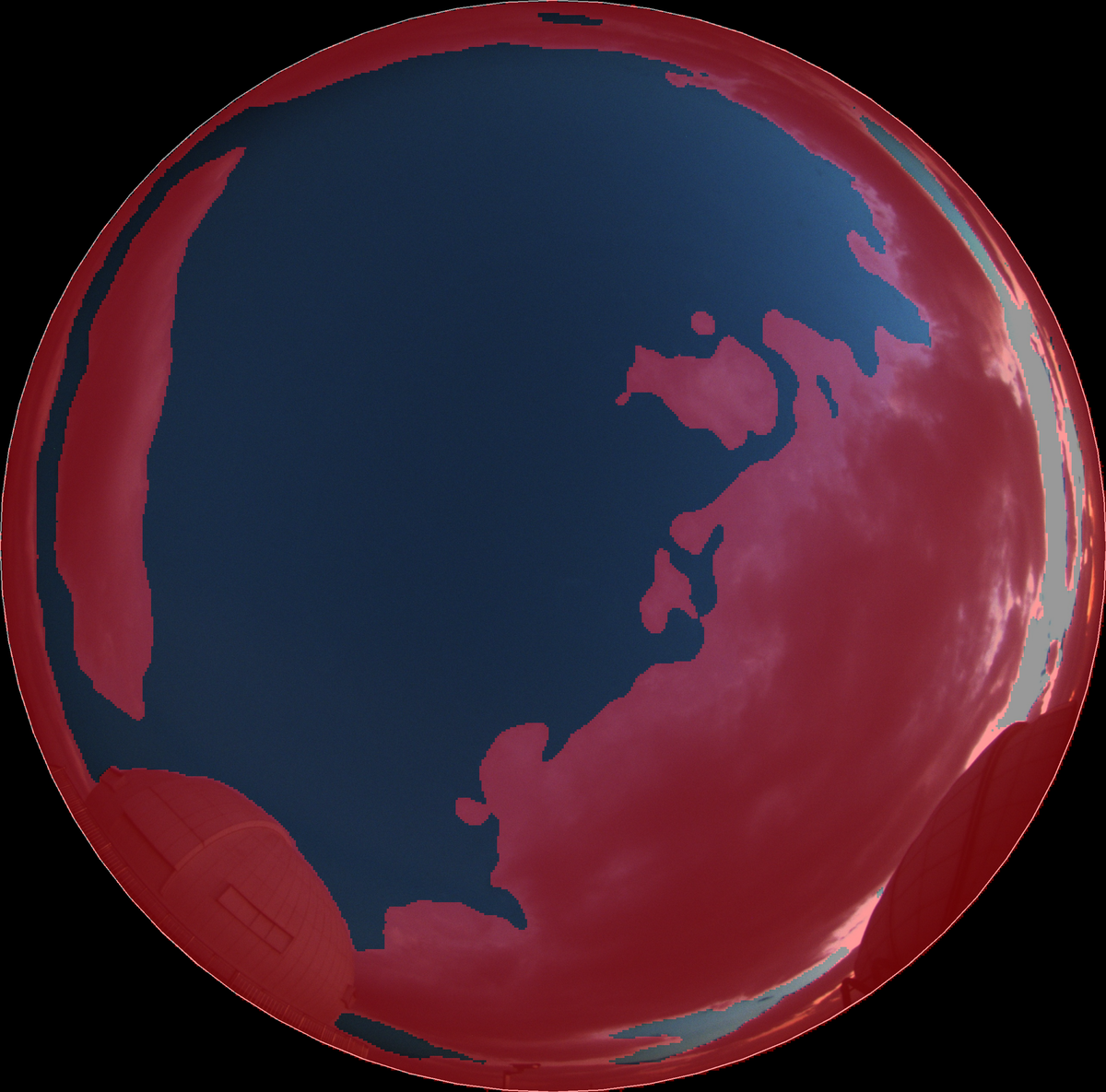} &  
        \includegraphics[width=0.9\linewidth]{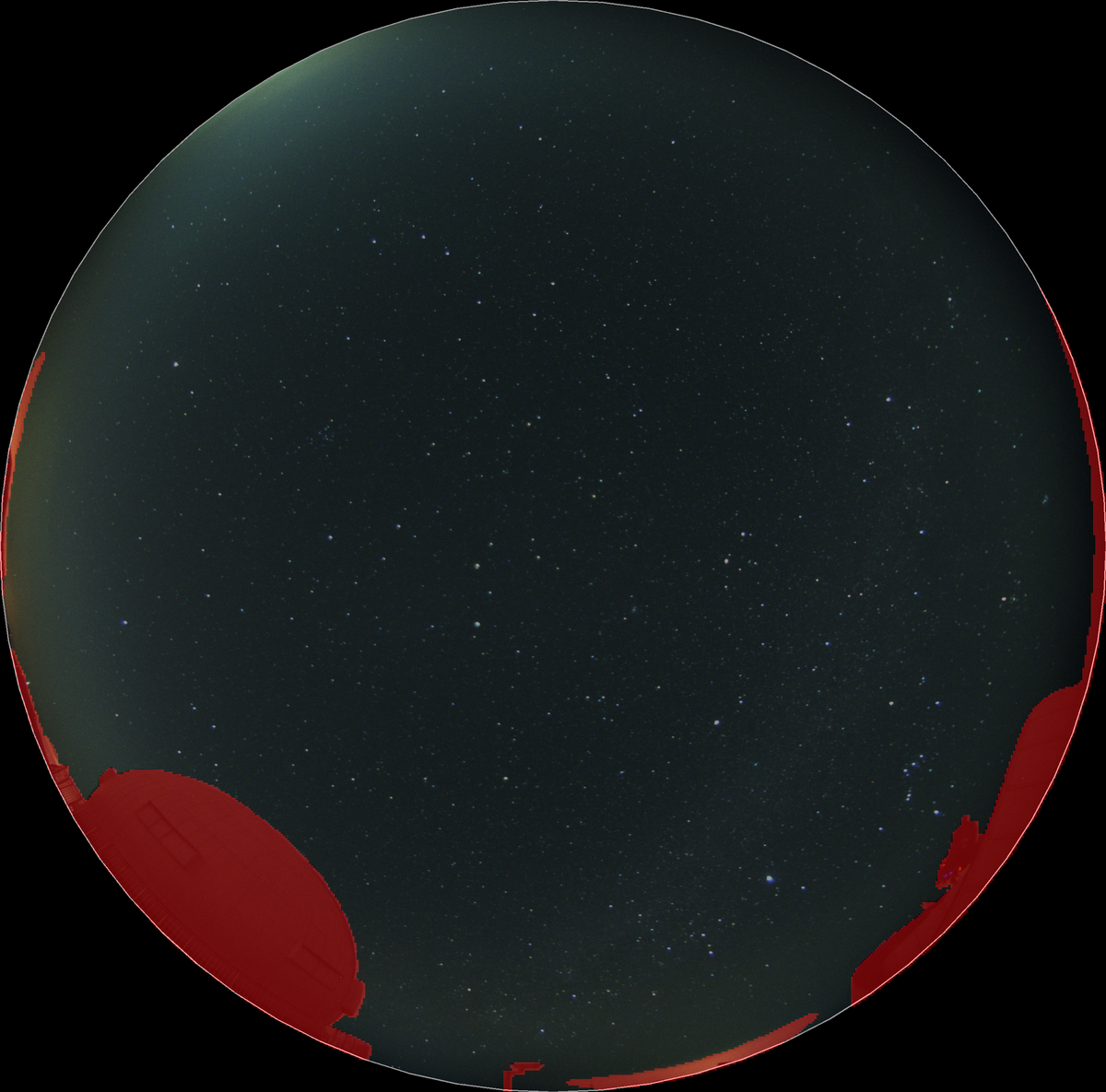} \\
        \midrule  
        \textbf{Difference Map} &  
        \includegraphics[width=0.9\linewidth]{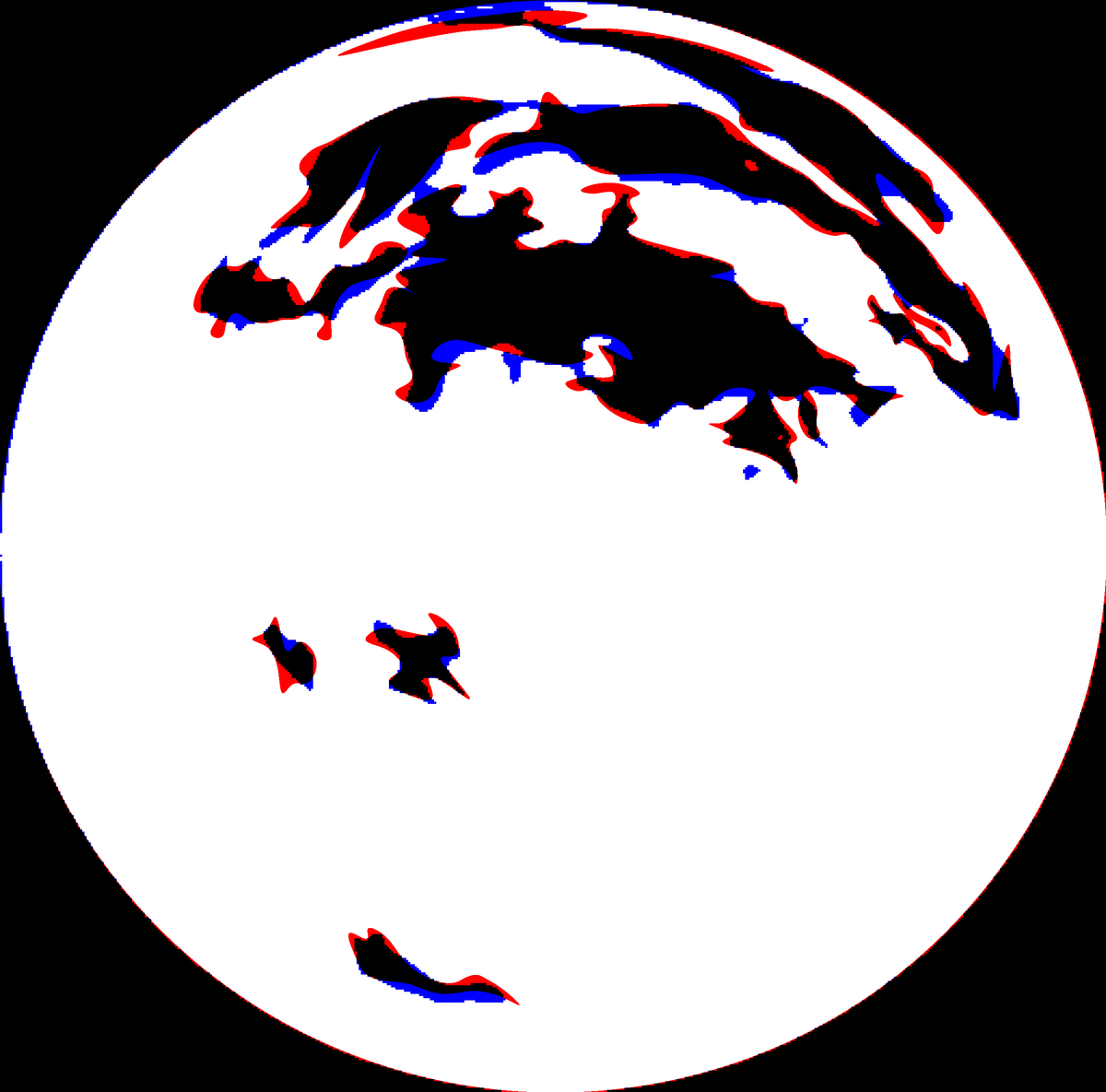} &  
        \includegraphics[width=0.9\linewidth]{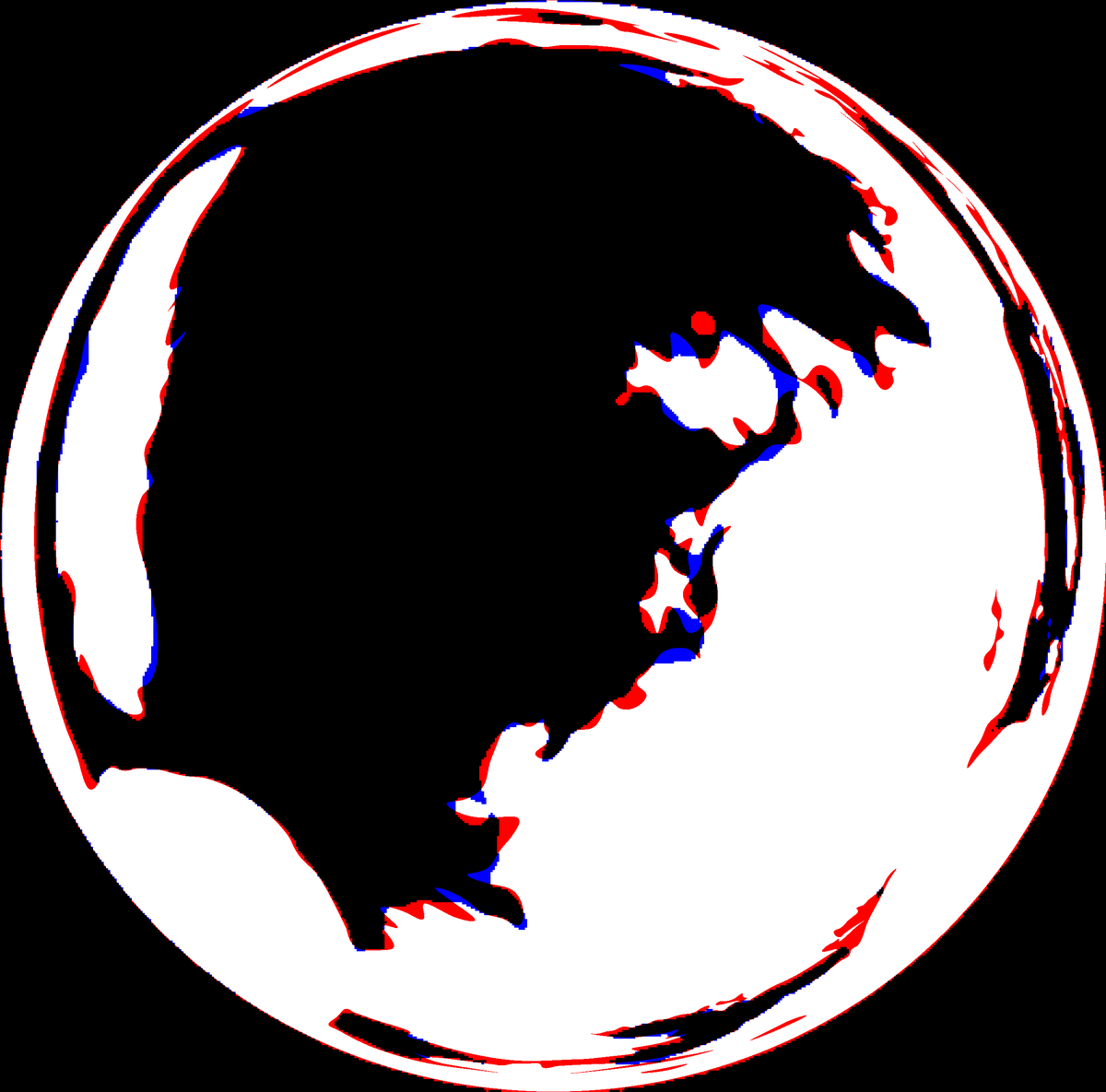} &  
        \includegraphics[width=0.9\linewidth]{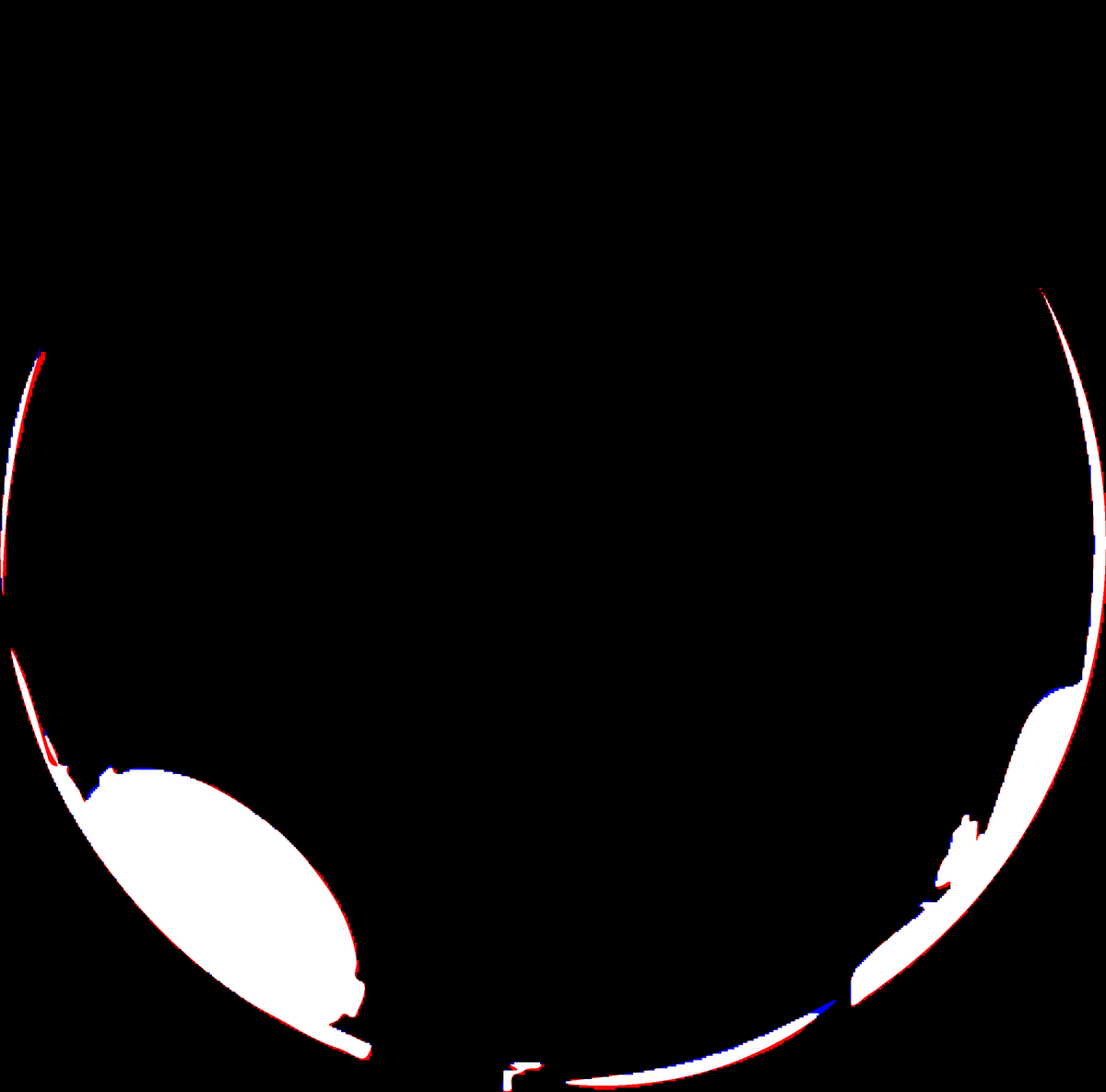} \\
        \midrule  
        \textbf{Difference Ratio} &  
        3.56\% & 3.96\% & 0.33\% \\
        \bottomrule  
    \end{tabular}  
    \caption{  
    Visualization of model performance for three categories (High, Moderate, and Low Unobservable Region). From top to bottom, the panels show: the raw sky image; the model prediction mask; the ground truth mask; the overlay image (showing the overlap of model prediction and ground truth on the raw image); and the difference map. The difference map highlights discrepancies between the model prediction and ground truth: red indicates areas incorrectly predicted as unobservable regions by the model (false positives), while blue indicates unobservable regions missed by the model (false negatives). The final row Difference Ratio shows the pixel-wise discrepancy between the model prediction and ground truth.  
    }
\label{fig:unobservable_region_comparison}
\end{figure*}  

Table~\ref{table:results} presents a comprehensive comparison between the U-Net baseline and several Enhanced UNet configurations with different loss weight settings (\(w_1, w_2, w_3\)). The evaluation metrics—Intersection over Union (IoU), Precision, Recall, and F1 Score—collectively reveal the segmentation performance and robustness of each model. The U-Net baseline serves as a reference, while all Enhanced UNet variants outperform it across all metrics, demonstrating the effectiveness of the proposed architectural and loss function improvements.  

Among the Enhanced UNet configurations, the combination \(w_1 = 0.4\), \(w_2 = 0.2\), \(w_3 = 0.4\) delivers the best results, achieving an IoU of 0.9212, Precision of 0.9564, Recall of 0.9602, and F1 Score of 0.9537. These improvements are especially significant for IoU and F1 Score, which are critical for accurate nighttime sky segmentation. Experiments with a single loss component (\(w_1=1.0, w_2=0.0, w_3=0.0\), etc.) yield lower performance, underlining the importance of a balanced hybrid loss. Fig.~\ref{fig:performance_evaluation} clearly visualizes these trends, highlighting the stable and substantial performance gains achieved with balanced loss settings.

Beyond quantitative metrics, Fig.~\ref{fig:unobservable_region_comparison} offers qualitative comparisons in three representative unobservable region scenarios (high, moderate, and low). For each, the raw input image, predicted mask, ground truth, overlay, and difference map are provided. The final row reports the pixel-wise difference ratio, directly reflecting segmentation error. Overlay images exhibit close alignment between predictions and ground truth, while the difference maps (red for false positives, blue for false negatives) provide insight into model errors. Notably, the Enhanced UNet achieves a minimal difference ratio of just 0.33\% in the low unobservable region case, and maintains high accuracy in more challenging scenarios.  

The experiments are conducted on a robust dataset of 1,800 training and 200 test images, ensuring statistical reliability. Compared to the baseline, the Enhanced UNet demonstrates substantial improvements—particularly for IoU and F1 Score—in both quantitative and qualitative evaluations. These results validate the effectiveness of the SCSE attention mechanism and hybrid loss in addressing nighttime sky segmentation challenges.   

In summary, the hybrid loss design consistently delivers superior segmentation performance and remarkable robustness. Intuitively, BCE loss ensures stable pixel-wise learning and enhances boundary precision; Dice loss addresses class imbalance prevalent in large unobservable regions; and IoU loss reinforces region-level consistency and connectivity. The synergy of these components enables the network to generalize effectively across diverse nighttime conditions. Ablation studies further confirm that hybridizing these losses is essential, as single-loss or pairwise combinations result in noticeably reduced accuracy.

\section{Applications of Cloud Coverage Detection in the Unattended Operation of Mephisto Telescope}  
\label{sec:coordinate_mapping}
In the previous section, the Enhanced UNet model is employed to identify unobservable regions. To incorporate the results into the Observation Control System (OCS), the pointing coordinates (RA, DEC) of the telescope during observation needs to convert to the pixel coordinates $(x, y)$ on the all-sky camera image. First, this section will introduce the modeling of the all-sky camera and its application in the telescope. Then we will establish a transformation system to link between telescope pointing coordinates and image pixel coordinates.

\subsection{Modeling of World Coordinates and Image Coordinates for All-sky Camera Image}  

Accurately determining whether a telescope’s pointing intersects unobservable regions, and achieving a precise transformation between celestial coordinates and image pixels, are essential yet nontrivial tasks in wide-field astronomical observations. Due to the geometric distortion inherent in all-sky camera systems, standard astrometric tools such as \texttt{astrometry.net} (\citealt{Lang+etal+2010}) are inadequate for directly deriving a reliable mapping from equatorial coordinates (RA, Dec) to the pixel coordinates $(x, y)$. Here, we adopt a two-stage transformation approach. First, equatorial coordinates are converted to horizontal coordinates—namely, altitude (Alt) and azimuth (Az)-based on the time and geographical location of observatory. Then, a Zenithal Equal-Area (ZEA) projection (\citealt{Calabretta+Greisen+2002}) is employed to project the (Alt, Az) pair onto the pixel coordinates. Compared to traditional tangent-plane (TAN) projections, the ZEA projection offers better performance in preserving areal relationships and handling the wide angular extent from zenith to horizon, making it particularly well-suited for all-sky imaging applications (\citealt{Xie+etal+2025}).

\subsubsection{Database Building}

\begin{figure*}[t]
\captionsetup[subfigure]{justification=centering} 
\centering  
\begin{subfigure}[b]{0.46\textwidth}  
    \centering  
    \includegraphics[height=6.0cm]{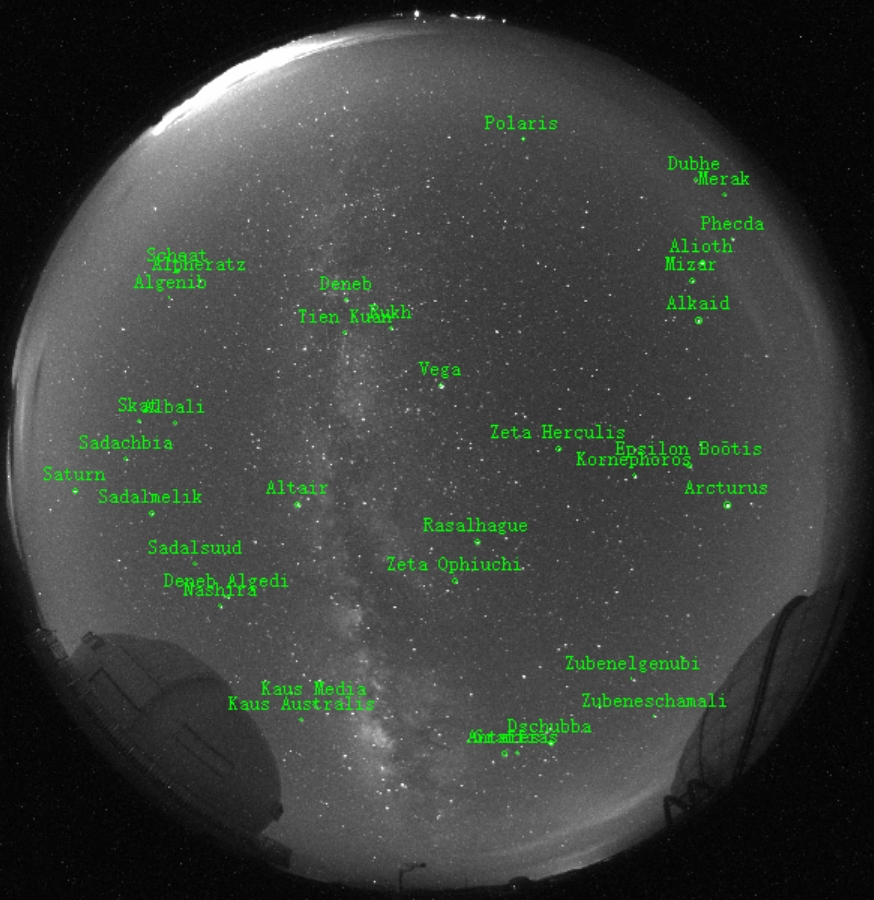}  
    \caption{Initial manually annotated dataset \\with sparse reference stars.}   
    \label{fig:dataset_rough}  
\end{subfigure}%
\hspace{0.01\textwidth}  
\begin{subfigure}[b]{0.46\textwidth}  
    \centering  
    \includegraphics[height=6.0cm]{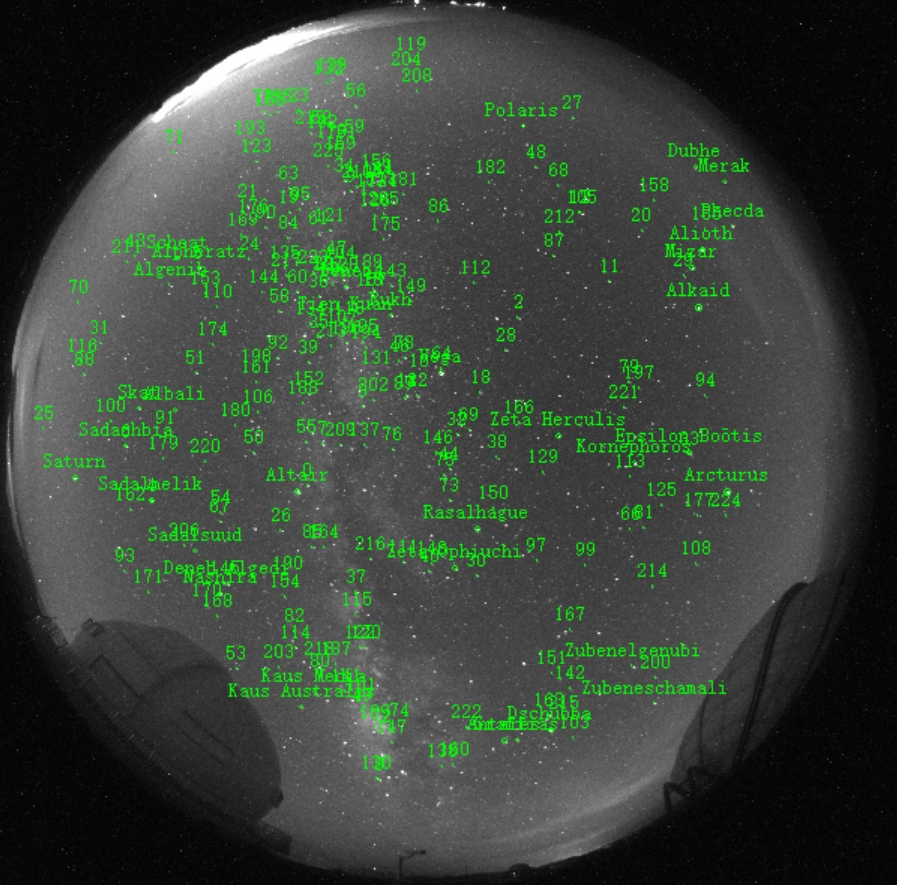}  
    \caption{Expanded dataset using Gaia DR3 \\and refined centroid matching.}   
    \label{fig:dataset_refined}  
\end{subfigure}  
\caption{Two-stage dataset construction for all-sky coordinate mapping. (a) An initial sparse WCS is built from manually selected stars spanning zenith to horizon. (b) A refined, dense calibration dataset is created by matching Gaia DR3 stars with local centroids, allowing high-precision model fitting.}  

\label{fig:dataset_constructing}  
\end{figure*}

To establish the mapping between horizontal coordinates and image pixel coordinates, we adopted a two-stage dataset construction strategy. As shown in Fig.~\ref{fig:dataset_rough}, we first selected a high-quality all-sky image captured under clear, moonless conditions and manually annotated a small number of bright stars from zenith to horizon. This initial sparse dataset allowed us to derive a coarse World Coordinate System (WCS) solution using the ZEA projection model.

Based on the initial calibration calibration, we retrieved all Gaia DR3 stars brighter than magnitude 8 within $\pm90^\circ$ zenith distance and pmra $<$ 20 and pmdec $<$ 20. Their (Alt, Az) coordinates are projected into preliminary $(x, y)$ image coordinates. We then used \texttt{Source Extractor} (\citealt{Bertin+Arnouts+1996}) to identify actual star positions in the image and applied nearest-neighbor or local optimization matching to refine the associations. The resulting dense set of matched star pairs, illustrated in Fig.~\ref{fig:dataset_refined}, enabled robust polynomial fitting and significantly improved model accuracy.

\subsubsection{Coordinate Transformation Model}

The coordinate transformation between celestial coordinates and pixel coordinates is realized through a two-stage approach: (1) projection from horizontal coordinates (Alt, Az) to intermediate planar coordinates $(\xi, \eta)$ via ZEA projection, and (2) polynomial mapping from $(\xi, \eta)$ to image pixel coordinates $(x, y)$.

The ZEA projection centers on the zenith and is well-suited for all-sky cameras with ultra-wide fields of view due to its area-preserving property. Given the $\mathrm{Alt}$ and $\mathrm{Az}$ of a celestial object, the zenith distance $c$ is computed as:
\begin{equation}
c = 90^\circ - \mathrm{Alt}
\label{eq:zenith_distance}
\end{equation}

Converting to radians:
\begin{equation}
c_{\mathrm{rad}} = \frac{\pi}{180} \cdot c, \quad \mathrm{az}_{\mathrm{rad}} = \frac{\pi}{180} \cdot \mathrm{Az}
\label{eq:rad_conversion}
\end{equation}

The ZEA projection equations are:
\begin{equation}
\begin{aligned}
\xi &= 2R \sin\left(\frac{c_{\mathrm{rad}}}{2}\right) \cos\left(\mathrm{az}_{\mathrm{rad}}\right) \\
\eta &= 2R \sin\left(\frac{c_{\mathrm{rad}}}{2}\right) \sin\left(\mathrm{az}_{\mathrm{rad}}\right)
\end{aligned}
\label{eq:zea_projection}
\end{equation}
Here, $R$ is the projection scale factor, typically set such that the projected radius matches the pixel radius from zenith to horizon, ensuring that $(\xi, \eta)$ are in pixel units.

To account for optical and projection distortions, we fit the intermediate coordinates $(\xi, \eta)$ to the final image coordinates $(x, y)$ using a two-dimensional polynomial of order $n$:
\begin{equation}
\begin{aligned}
x &= \sum_{i+j \leq n} a_{ij} \xi^i \eta^j, \\
y &= \sum_{i+j \leq n} b_{ij} \xi^i \eta^j
\end{aligned}
\label{eq:poly_mapping}
\end{equation}

For $n = 2$, this yields a total of 6 terms for $x$ and 6 terms for $y$, resulting in 12 coefficients to be determined. These include constant, linear, and quadratic components, enabling the model to capture nonlinear distortions introduced by the camera optics and the projection process.
\subsubsection{Residual Analysis}

To evaluate the accuracy of the coordinate transformation model, we calculate the residual error for each calibration star as the Euclidean distance between its predicted and measured image coordinates:
\begin{equation}
\mathrm{residual} = \sqrt{\left(x_{\mathrm{pred}} - x_{\mathrm{meas}}\right)^2 + \left(y_{\mathrm{pred}} - y_{\mathrm{meas}}\right)^2}
\label{eq:residual}
\end{equation}
where $(x_{\mathrm{meas}}, y_{\mathrm{meas}})$ are the centroid positions extracted from the all-sky image, and $(x_{\mathrm{pred}}, y_{\mathrm{pred}})$ are the corresponding predicted positions from the transformation model.

The mean residual across all matched stars quantifies the overall accuracy of the geometric model. Smaller residuals indicate improved correspondence between the projected sky model and actual stellar positions in the image. 

\subsubsection{Model Fitting and Discussion}

\begin{figure}[t]
  \centering
  \begin{subfigure}[t]{\textwidth}
    \centering
    \includegraphics[height=6.0cm, keepaspectratio]{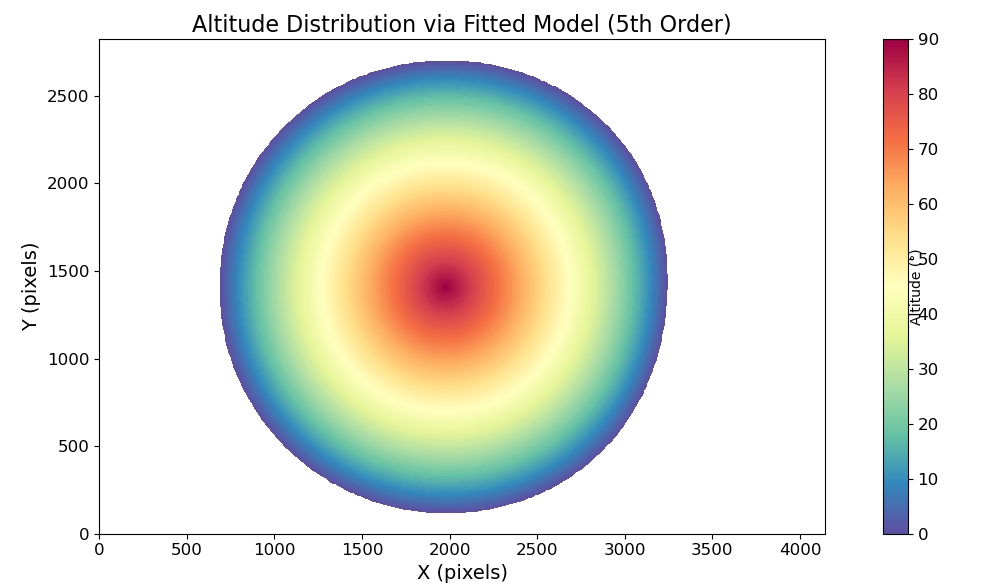}
    \caption{Modeled ALT distribution in image coordinates.}
    \label{fig:top}
  \end{subfigure}
  \vspace{\bigskipamount} 
  \begin{subfigure}[t]{\textwidth}
    \centering
    \includegraphics[height=6.0cm, keepaspectratio]{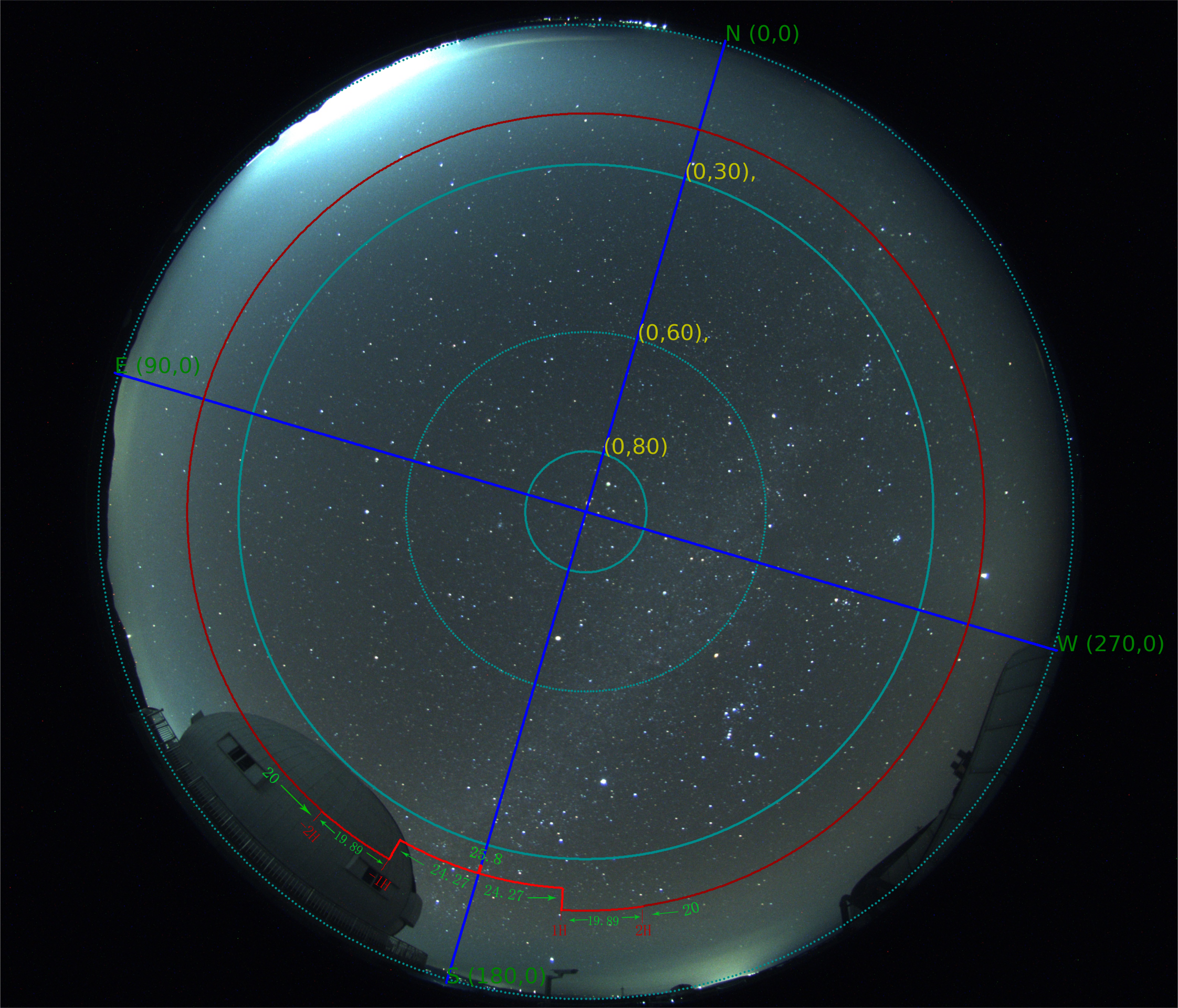}
    \caption{Actual all-sky image with ALT/AZ overlay and pointing limit.}
    \label{fig:bottom}
  \end{subfigure}
  \caption{Coordinate transformation evaluation using ZEA projection.  
  (a) Symmetric ALT gradient centered on zenith; 
  (b) Geometric alignment between model and stellar positions. The red curve marks the lower limit of the Mephisto telescope pointing.}
  \label{fig:modelfitsandreal}
\end{figure}

To model the celestial-to-pixel coordinate transformation, we implement a fifth-order polynomial ($n = 5$) for the transformation described in Eq.~\eqref{eq:poly_mapping}. This yields a mean residual of 0.95 pixels, corresponding to an image scale of 0.0621 degrees per pixel. To evaluate the quality of the transformation, we generate an ALT distribution map across the image plane, where pixel colors represent modeled ALT angles. As shown in Fig.~\ref{fig:modelfitsandreal}, the top panel illustrates the fitted ALT field. The resulting smooth and radially symmetric gradient confirms that the polynomial mapping preserves the sky's geometric structure. The bottom panel presents a real all-sky image under clear sky conditions, overlaid with the calculated ALT contours and AZ reference lines derived from the transformation. Concentric circles correspond to ALT levels (e.g., 30°, 60°, and 80°), while cardinal directions (N, E, S, W) are labeled in green. A red curve indicates the practical ALT limit of the Mephisto telescope, which defines the lower boundary for reliable observations due to mechanical constraints. The excellent agreement between the modeled ALT contours and the observed stellar background demonstrates the effectiveness and low distortion of the fitted projection model, validating its use in visibility estimation and dynamic scheduling.

Building upon the Enhanced UNet segmentation model (Section \ref{sec:methodology}) that identifies unobservable regions and the coordinate transformation pipeline (Section \ref{sec:coordinate_mapping}) using ZEA projection to map celestial coordinates to image pixels, this system establishes a coordinate mapping that transcends geometric alignment. It enables real-time, pixel-level visibility queries for any telescope pointing (defined by RA/DEC): by converting the pointing to its corresponding location in the all-sky image and ultimately the ALT-AZ system, the system instantly determines if it intersects a segmented obscured area (e.g., cloud cover). Furthermore, a configurable altitude buffer (e.g., 5–10°) prevents scheduling targets near the image edge – where observing conditions degrade rapidly – significantly increasing scheduler resilience.

\begin{figure}[t]
  \captionsetup[subfigure]{justification=centering} 
  \centering
  \begin{subfigure}[b]{0.46\textwidth}
    \centering
    \includegraphics[height=6.0cm]{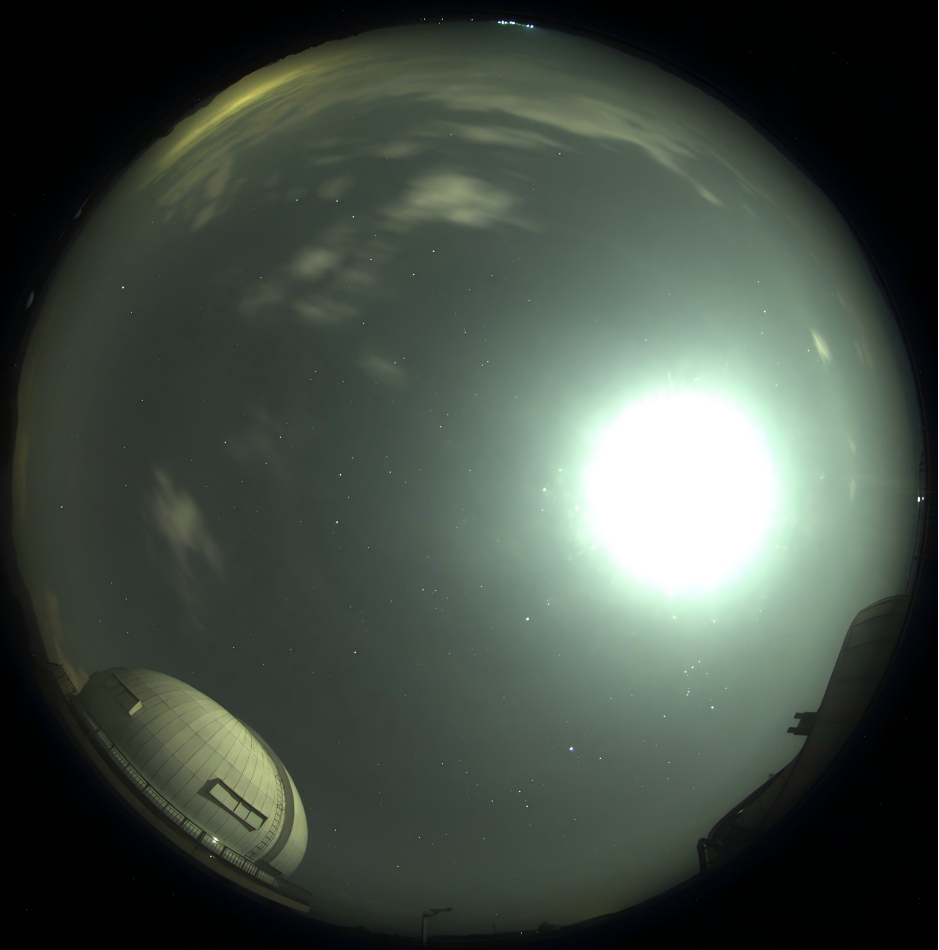}
    \caption{Raw all-sky image captured at Lijiang Observatory under strong moonlight conditions.}
    \label{fig:va_left}
  \end{subfigure}
  \hspace{0.01\textwidth}
  \begin{subfigure}[b]{0.46\textwidth}
    \centering
    \includegraphics[height=6.0cm]{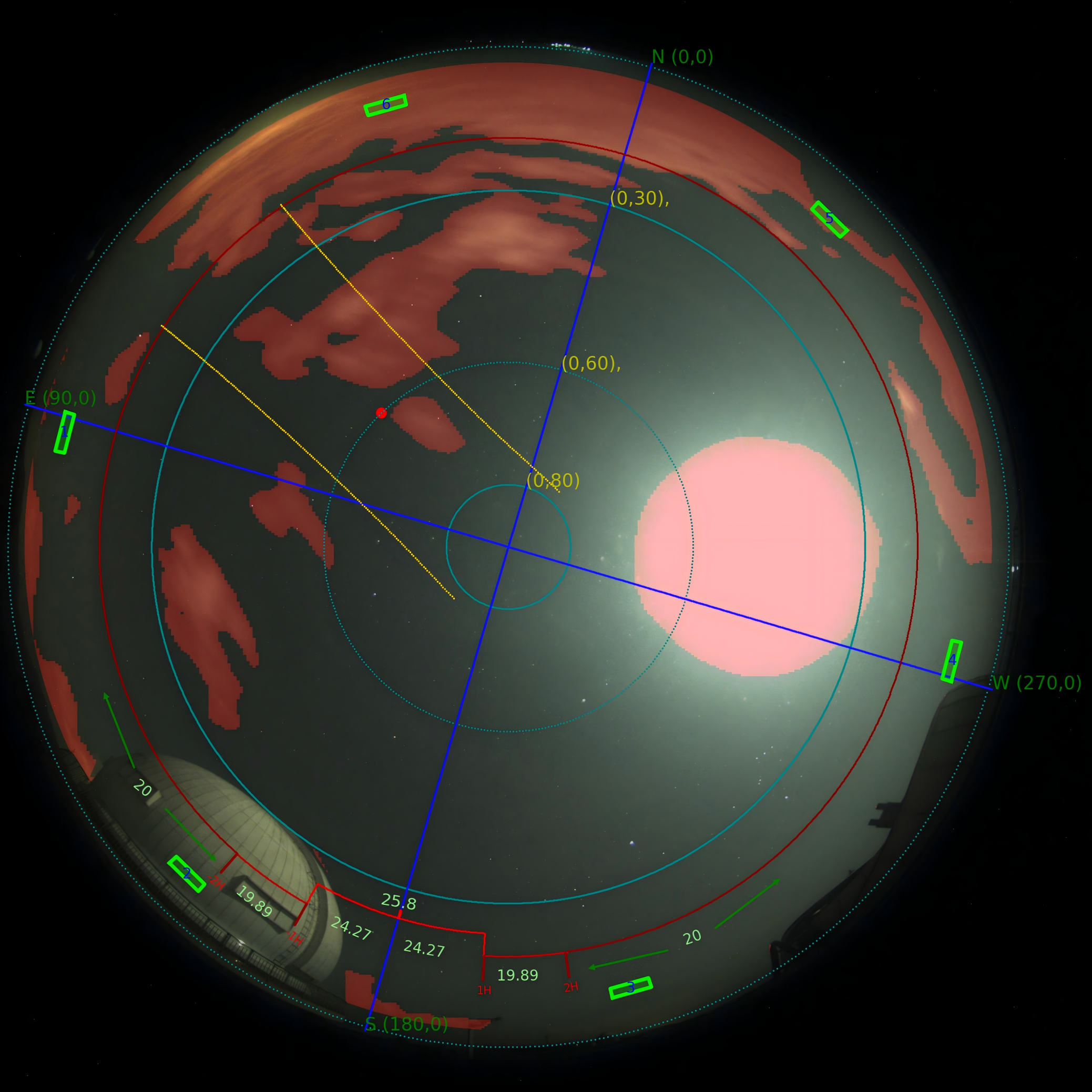}
    \caption{Segmentation mask (red) indicating cloud/moon-obscured areas.}
    \label{fig:va_right}
  \end{subfigure}
  \caption{Example of cloud-aware scheduling based on segmentation and coordinate mapping. The telescope’s pointing (red dot) lies near a cloud boundary.}
  \label{fig:va_results}
\end{figure}

The system's capability goes beyond binary visibility judgment (whether the target is observable). The output of the segmentation model supports quantitative evaluation of sky quality: by calculating the proportion of unobservable pixels relative to the total projected area of the celestial dome (whether caused by clouds, moonlight scattering, or other atmospheric effects), cloud coverage can be obtained in real-time as a percentage. This type of statistics is crucial for scheduling systems, as it can be used to prioritize clear sky areas, reweight target queues, or evaluate the feasibility of observations throughout the night. Although moonlight scattering may sometimes exaggerate cloud coverage estimates, this "pollution" itself corresponds to actual decreasing photometric degradation, and therefore has guiding significance for sky survey operations. This capability also helps alleviate the limitations of infrared cloud sensors (IRCS), which may provide misleading readings when encountering high humidity or moonlight during unmanned operation. This function form the basis for visibility aware scheduling. It will be implemented into the Mephisto Observation Control System (OCS) in future. An example demonstrating how the segmentation and coordinate mapping pipeline supports cloud-aware scheduling is shown in Fig.~\ref{fig:va_results}. Once deployed, OCS will automatically compare telescope pointing with segmentation mask, avoid unobservable areas (or close to an angular distance buffer around it), and dynamically rearrange the observation.

\begin{table}[t]
\bc
\begin{minipage}[]{100mm}
\caption[]{Comparison of Nighttime Sky Visibility estimation Methods \label{tab:methods-comparison}}
\end{minipage}
\setlength{\tabcolsep}{2pt}
\small
\begin{tabular}{cccc}
\noalign{\smallskip}\hline\noalign{\smallskip}
\textbf{Feature} & \textbf{Photometry/Extinction} & \textbf{Photometry-Free (PFSAVE)} & \textbf{ Segmentation-based (This work)} \\
\noalign{\smallskip}\hline\noalign{\smallskip}
Input          & FITS + catalog & FITS + catalog &  All-sky images \\
Granularity    & Grid (1--2$^\circ$) & Grid (1--2$^\circ$) & Pixel-level ($<$0.1$^\circ$) \\
Latency        & Seconds--minutes & Seconds & Milliseconds \\
Moon/Cloud Robust & Poor & Moderate & good \\
Interpretability & Physical extinction (mag) & Limiting mag per grid & Pixel mask, supports aggregation \\
Accuracy       & $\sim$0.05 mag (clear) & $r>0.85$ (vs. expert) & IoU 0.92, F1 0.95 (pixel)\\
Primary Application & Sky condition statistics & Dynamic grid scheduling & Real-time fine-grained scheduling \\
\noalign{\smallskip}\hline
\end{tabular}
\ec
\end{table}

We compare our method with others. Photometry/extinction-based methods (\citealt{zhi+etal+2024}) rely on extracting stellar sources and comparing their measured fluxes to catalog values within spatial grids (such as HEALPix). This framework provides physically interpretable extinction and transparency estimates, and its results are often benchmarked by quantitative agreement with manual ratings (e.g., median error of $\sim$0.05 mag in clear conditions). However, these methods are sensitive to the quality of source extraction and accurate catalog matching. Performance is readily influenced by strong moonlight, sky gradients, incomplete calibration, and photometric outliers. Their spatial granularity is set by the chosen grid size (e.g., typically $1^{\circ}$–$2^{\circ}$ per cell), and the computation introduces non-negligible latency due to repeated photometric analysis, which may not meet the requirements of low-latency, real-time scheduling for robotic observations.

Photometry-free algorithms, such as the PFSAVE method (\citealt{Xie+etal+2025}), infer visibility in each grid cell based not on photometric calibration, but on the limiting magnitude of robustly detected stars. This approach incorporates precise astrometric transformations and carefully designed detection metrics (e.g., S/N thresholds, T-metrics) to minimize false positives from noise, clouds, or moonlight. Although PFSAVE can nominally process lower-quality images, high-precision visibility estimation still requires FITS data and reliable star matching. The resulting grid-wise visibility maps correlate well with expert assessments (e.g., Spearman $r > 0.85$ under cloud cover). However, its spatial resolution is fundamentally limited by the grid definition, and real-time performance is affected by the computational demands of star extraction and assignment.

In contrast, our segmentation-based approach adopts an end-to-end deep learning framework. It directly produces pixel-level visibility masks from standard all-sky images, eliminating the need for specific FITS photometry, stellar extraction, or external catalogs. This enables real-time, high-resolution sky condition mapping robust to faint clouds, cloud edges, and complex illumination, including scattered moonlight. Empirical experiments demonstrate that our method achieves millisecond-level inference per frame on modern GPUs, supporting true online operation in unattended environments. Furthermore, segmentation masks can be aggregated into arbitrary grid scales if needed, enabling both fine-grained decision-making and seamless integration with grid-based pipelines.The main properties and performance of the three methods are presented in Table 4.

\section{Conclusion}  
\label{sec:conclusion}  

In this study, we have developed a deep learning-based framework for the segmentation and mapping of unobservable regions---such as clouds and moon-scattered areas---in nighttime all-sky images, with the goal of enhancing the efficiency and robustness of autonomous astronomical observations. Specifically, we designed an enhanced UNet architecture that incorporates an EfficientNet-B4 encoder, SCSE attention modules in the decoder, and a compound loss function combining Binary Cross-Entropy, Dice, and IoU losses. Extensive experiments conducted on a manually annotated dataset of 2,000 nighttime all-sky images demonstrate that our model achieves state-of-the-art segmentation performance. With the optimal loss weight configuration ($w_1=0.4$, $w_2=0.2$, $w_3=0.4$), the model attains an Intersection over Union (IoU) of 0.9212, Precision of 0.9564, Recall of 0.9602, and F1 Score of 0.9537 on the test set. Qualitative analysis further reveals that the pixel-wise difference between predicted and ground-truth masks is as low as 0.33\% under clear-sky conditions, and remains below 4\% even in challenging cases with substantial cloud coverage.  

To enable practical deployment in robotic observatories, we have established a Zenithal Equal Area (ZEA)-based coordinate transformation pipeline, which accurately maps telescope pointing coordinates (RA, Dec) to all-sky image pixels. The transformation model, fitted with a fifth-order polynomial, achieves a mean residual of just 0.95 pixels, ensuring highly precise alignment between celestial and image coordinates. This system supports real-time, pixel-level visibility checks for any telescope pointing, and provides quantitative cloud coverage statistics to inform dynamic scheduling decisions. The framework is planned to be integrated into the Mephisto telescope's observation control system (OCS), thereby enabling cloud-aware, fully autonomous sky survey operations.  

Looking ahead, several directions remain for further improvement. To better capture dynamic weather patterns, the development of a temporal model for cloud evolution could enable short-term predictions of sky conditions. In addition, further refinement of the segmentation model will be pursued to address photometrically challenging scenarios, such as extremely faint clouds or strong moonlight contamination. Seamless integration of the segmentation and coordinate mapping pipeline with the observatory control system will also be a focus of future work, supporting real-time, weather-adaptive scheduling and robust data quality control. Finally, expanding the annotated dataset and exploring transfer learning strategies may further enhance the generalization capability of the framework to other observatories and diverse atmospheric environments.

In summary, the developed deep learning framework enables precise segmentation of unobservable sky regions and accurate mapping between celestial and image coordinates, greatly enhancing the autonomy and adaptability of astronomical observations. With its planned integration into the Mephisto observation control system, the method lays a solid foundation for real-time, weather-aware sky surveys. Looking forward, expanding the dataset and introducing temporal modeling will further broaden the applicability and robustness of this framework across diverse observational environments.

\normalem
\begin{acknowledgements}

This work is supported by the National Key R\&D Program of China (Grant No. 2024YFA1611600), the Yunnan Fundamental Research Projects (Grant No. 202501AT070455), and the Yunnan Revitalization Talent Support Program, including the Science \& Technology Champion Project (Grant No. 202005AB160002), the Innovation Team Project (Grant No. 202105AE160021), and the Top Team Project (Grant No. 202305AT350002). The authors also acknowledge support from the Key Laboratory of Survey Science of Yunnan Province with Project No. 202449CE340002. We would like to thank the staff of the Lijiang Observatory for their assistance with data acquisition and system maintenance. Additionally, we are grateful to the members of the Mephisto project team for valuable discussions and technical assistance. Special thanks are extended to the anonymous reviewers for the constructive comments and suggestions, which have greatly improved the quality of this manuscript. We also acknowledge the use of the Gaia DR3 catalog and the open-source software tools utilized in this research.

\end{acknowledgements}

\bibliographystyle{raa}
\bibliography{bibtex}

\end{document}